# Exploiting Correlations for Expensive Predicate Evaluation

Manas Joglekar
Stanford
manasrj@cs.stanford.edu

Hector Garcia-Molina
Stanford
hector@cs.stanford.edu

Aditya Parameswaran
U. Illinois and MIT
adityagp@illinois.edu

Christopher Re
Stanford
chrismre@cs.stanford.edu

## ABSTRACT

User Defined Function(UDFs) are used increasingly to augment query languages with extra, application dependent functionality. Selection queries involving UDF predicates tend to be expensive, either in terms of monetary cost or latency. In this paper, we study ways to efficiently evaluate selection queries with UDF predicates. We provide a family of techniques for processing queries at low cost while satisfying user-specified precision and recall constraints. Our techniques are applicable to a variety of scenarios including when selection probabilities of tuples are available beforehand, when this information is available but noisy, or when no such prior information is available. We also generalize our techniques to more complex queries. Finally, we test our techniques on real datasets, and show that they achieve significant savings in cost of up to 80%, while incurring only a small reduction in accuracy.

## 1. INTRODUCTION

User defined functions (UDFs) provide query languages with extra, application-dependent functionality. UDFs are especially important in data science, enabling data scientists to augment their workflows with complex computation, including calls to machine learning algorithms or external APIs. Overall, UDFs come in many flavors: they could invoke external services (e.g., calls to an up-to-date weather monitoring service, or a credit check service), expensive algorithms (e.g., image, video or text analysis algorithms), or even crowdsourced workers (e.g., workers on Mechanical Turk in CrowdDB or Deco [17, 33]). UDFs have become even more common in the past few years, due to the increasing use of crowdsourcing [17, 33], large scale machine learning [22, 32], and scientific procedures [11] in conjunction with traditional database systems. In all of these cases, UDFs are costly to evaluate, either (i) in terms of time, for instance in the expensive algorithm scenario, or (ii) in terms of monetary cost, for instance in the crowdsourcing or external service call scenario.

In this paper we study techniques for reducing the number of UDF invocations during execution of a select query with a UDF predicate, provided the user tolerates some reduction in accuracy. In our case we define accuracy in information retrieval terms (and not in terms of errors of individual values). That is, say $C$ (a set of tuples) is the correct result of a query (performing all UDF calls), and $R$ is the approximate result. Then the precision is $|R \cap C|/|R|$ and the recall is $|R \cap C|/|C|$. The user specifies bounds on precision and recall to indicate what accuracy is acceptable.

EXAMPLE 1.1. *To motivate our approach, consider the following simple example. A user wants to contact customers with a very good credit rating to offer a special promotion. The following query describes his needs:*

Q : SELECT ∗ FROM R(A, ID) WHERE f(ID) = 1.

*Here* f *involves a call to a UDF, say a credit check service, and returns 1 (for "good") or 0 ("bad") depending on whether or not the credit score is above a threshold.*

*If the user requires perfect precision and recall, then the system must retrieve every* R *record and check if its* f(ID) *value is* 1. *The cost, either in terms of time taken or money paid to the credit bureau, will be high. Instead say the user is willing to tolerate a slightly lower value of precision and/or recall, say 90% for each. (The precise meaning of this bound will be discussed later.) In other words, the user is willing to pitch the promotion to a few customers with a bad credit rating, and to miss a few good customers, in order to get the query results significantly faster or cheaper.*

*Then, one straightforward query execution strategy is to simply evaluate* f *on as many tuples as necessary in sequence until the recall and precision constraints are met. Thus, we will end up making close to $90\% \times 3000 = 2700$ calls to the credit check* f *function.*

*Another approach, which we will explore in depth in this paper, is to estimate and then exploit correlations. In our example, let us assume there is an attribute* A *of* R *that is correlated to the output of* f. *For instance,* A *could be an income attribute or a housing status that could indicate a higher likelihood for having a good credit score. (We discuss later how we can discover one or more attributes that may be correlated to the UDF.)*

**Step 1: Estimating Correlations.** *One way to estimate correlations is by sampling some* R *tuples. For example, after reading 5% of the tuples with each* A *value, the system has the following estimates:*

- *If* A=1, f(ID) = 1 *with probability 0.9*
- *If* A=2, f(ID) = 1 *with probability 0.5*
- *If* A=3, f(ID) = 1 *with probability 0.1*

*(As we discuss later, sampling is not the only way to obtain the statistics needed by our next step.) The system also knows that there are 1000 tuples with each of the possible* A *values, for a total of 3000* R *records.*

**Step 2: Executing Query.** *Then, a better query execution strategy is to handle tuples with each* A *value differently: Tuples with* A=1 *are very likely to satisfy the predicate, so we can add them to* *the result without evaluating* f. *Tuples with* A = 3 *are unlikely to match, so we can simply discard them. Tuples with* A = 2 *can go either way, so we evaluate* f *on each tuple with* A = 2 *and only add the tuples that match to the result. With this strategy we only need to evaluate* f *1000 times (plus the evaluations needed for sampling), not the 2700 calls of the straightforward approach, and we still get an expected precision and recall of (approximately) 90%.*

*Note that the execution strategy illustrated above is one of many: for every group of tuples that shares the same* A *value, we can choose to (a) directly add all of them to the result without evaluating* f, *(b) discard all of them, or (c) evaluate all of them and then decide whether to add it to the result or not. Thus, if* n *is the number of distinct values of* A, *we have* $3^n$ *possible execution strategies. We could also employ probabilistic execution strategies, where we toss a biased coin for each tuple in a group of tuples sharing the same* A *value, and decide what to do based on the outcome. Overall, it is not immediately obvious which of these strategies we should use in order to meet our precision and recall guarantees. Indeed, as we will see in the following, designing optimized execution strategies is NP-Hard.*

Our approach avoids UDF evaluations by exploiting *correlations* between the results of boolean UDF calls and values of other categorical attributes. In our example we have swept a number of details and possible generalizations "under the rug" in order to present the essence of our approach. The details will be formalized in the sections that follow, but we make a few clarifications:

- The attribute A in our example could be several attributes or even a virtual attribute representing the output of a machine learning algorithm that predicts the value of f based on values in a tuple. That is, with appropriate training data we may develop a less expensive function f' that predicts f (and also provides confidences for its predictions). With such a function we can skip Step 1, and in Step 2 use the f' values to decide whether to discard a tuple, output it immediately, or evaluate the expensive f function. We will focus on the sampling approach because it makes it easier to account for the cost of obtaining correlations.

  That said, our techniques for step 2 are agnostic to the source of the confidences values. As a result, we can use confidences generated by say, logistic regression or least squares estimation, or any other statistical techniques. We experimentally evaluate the impact of where the confidences come from (including sampling and logistic regression) in Section 6.

- We assume that f values have not been cached in advance. If the values are available, then the expense of evaluating f on every tuple has been paid in advance, and conventional query processing strategies can be used. Our techniques can be applied even if some of the values are available in advance. Note that in many cases f outputs vary with time and should not be cached. For example, weather predictions or credit ratings change over time.

We reiterate that our approach hinges on two assumptions: (a) that the user is willing to accept imperfect precision and/or recall, and (b) that there is an attribute (possibly virtual, possibly multiple attributes) that is correlated to the UDF output. For (a), we believe that users are familiar with document retrieval models where it is not practical to get the complete and exact answer to a query. Often, the output of the UDF itself is subjective or approximate (e.g., whether a patient is prone to a particular illness, or an image is inappropriate), so errors cannot be ruled out no matter what we do. For (b), if correlated attributes are not known in advance, there are well known techniques for learning what attributes are good predictors, and for combining these attributes into a virtual attribute that can be used by our solution.

When these assumptions hold, we will show (with real data sets) that queries can be evaluated very efficiently. These savings will grow in importance as data sets grow in size and UDFs become more and more popular. In our experiments we will also study various performance related questions: For example, how much sampling should one do in Step 1? As we sample more, our statistics improve, but our cost savings decrease. How sensitive is our approach to the number of A values? As the number of values grows, the number of tuples with a given value shrinks, possibly making it harder to get estimates.

Prior work has addressed the optimal placement of UDF evaluation in query plans [14, 15, 20, 23], assuming users have provided hints for costs of UDFs; in our work, we aim to avoid UDF evaluations entirely by exploiting correlations. Our work is also similar to the work on exploiting correlations between sensors to reduce evaluations in the sensor networks domain [16], however, the scenario is very different, necessitating different approaches. There has been significant work on Approximate Query Processing [13, 18, 19, 21, 27]. However, these papers only approximate numerical aggregate quantities, while we focus on approximating set-valued queries. Finally, there is some recent work on finding correlations between attributes of a relation [12, 25, 28], which can then be used to better estimate costs of intermediate results for query optimization. Our work, on the other hand, exploits correlations between attributes in a relation and an external UDF. Related work is surveyed in more detail in Section 7.

**Contributions:** We propose a novel approach to optimize expensive UDFs by exploiting correlation information. Specifically,

- In Section 2, we introduce the relevant notation for the setup of the problems we consider.
- In Section 3, we study the case where correlations are known. We consider three scenarios where we can exploit existing correlations, and we show NP-Hardness for the first scenario and describe asymptotically optimal algorithms for the other two scenarios.
- In Section 4, we study the case where correlations are not known in advance, and we must jointly estimate correlations, and exploit them. We describe a method of obtaining correlation information, how to use this information adaptively, and provide a rule of thumb to use in practice.
- In Section 6, we experimentally evaluate the performance of our algorithms on four real datasets, and also demonstrate the correctness of the algorithms.

In Section 5, we extend our methods to more advanced SQL queries that involve more than one table and/or UDF predicate.

## 2. PRELIMINARIES

We begin by considering queries like our example query Q in Section 1. We reproduce the query here:

Q : SELECT * FROM R(A, ID) WHERE f(ID) = 1

The relation R contains an attribute A, the values of which are correlated with the output of f(ID) across tuples. It is straightforward to generalize to queries with more correlated attributes, with joins or projects, or with categorical predicates for f with operators other than equality. (We discuss generalizations in Section 5.)

In practical applications, the value of the correlated column A may not be known in advance. In Section 4.4, we describe a general procedure for finding the best correlated column in any table.

| Tuple No. | A | ID | f(ID) |
|---|---|---|---|
| 1 | 1 | 999-999-999 | 1 |
| 2 | 1 | 913-418-777 | 1 |
| 3 | 1 | 719-334-111 | 1 |
| 4 | 1 | 999-999-999 | 1 |
| 5 | 2 | 913-418-737 | 0 |
| 6 | 2 | 719-334-113 | 1 |
| 7 | 2 | 999-999-299 | 0 |
| 8 | 3 | 913-418-737 | 0 |
| 9 | 3 | 719-334-121 | 0 |
| 10 | 3 | 999-999-959 | 0 |
| 11 | 3 | 913-418-727 | 0 |
| 12 | 3 | 719-334-311 | 1 |

*Table 1: Example Data for* R

To assist the description of our terminology, we provide example data for R in Table 1. Naturally, f(ID) is not known in advance (and instead must be computed using UDF evaluations), but is also shown along with the table. The tuples in R that satisfy the predicate (i.e., f(ID) equals 1) are called *correct*, while those that do not are called *incorrect*. Thus, in Table 1, tuples 1-4, 6, and 12 are correct, while the rest are incorrect.

**Groups of Tuples:** We use A to denote both the correlated attribute, as well as the set of distinct values that appear in R.A. Therefore, we use $a \in A$ to denote a value that attribute A can take. In Table 1, $a$ can be 1, 2, or 3. The set of tuples that share the same value of A is called a *group*. We let $t_a$ be the number of tuples in the group that has A=$a$. Thus, $t_1 = 4, t_2 = 3$, and $t_3 = 5$ in our example. In a group corresponding to A=$a$, we let $c_a$ denote the number of correct tuples (i.e., tuples satisfying the predicate), and $w_a$ be the number of incorrect tuples. Thus, for $a = 2$, $c_2 = 1$, while $w_2 = 2$. While we assume $c_a, w_a$ to be known in our example (and also in our initial setting in Section 3.1), typically these values are not known, and are treated as random variables. In such cases, we use upper case $C_a$ and $W_a$ to denote the random variables. We assume $t_a$ for all $a \in A$ is always known.

**Actions and Costs:** For our simple query Q, we must decide if each tuple in R is in the result. We have three alternative *actions* we could take:
- First, we could *discard* the tuple, i.e., no action is taken on the tuple, and the tuple does not contribute to the output. In this case, we are predicting that the tuple is not correct.
- Next, we could *retrieve but not evaluate* the tuple, i.e., retrieve the tuple from R, and add it to the result without actually evaluating the UDF. In this case, we are predicting that the tuple is correct.
- Last, we could *retrieve and evaluate* the tuple, i.e., retrieve the tuple from R, evaluate the UDF on the tuple, and add the tuple to the result if it satisfies the condition f(ID)=1. Then we are certain to be accurate in our assessment of the tuple.

Say we incur a cost of $o_e$ for every tuple evaluated, and a cost of $o_r$ for every tuple retrieved from storage. Since UDFs are expensive, $o_e$ is likely to be much greater than $o_r$. Thus, the cost of discarding a tuple is 0, the cost of retrieving and not evaluating is $o_r$, and the cost of retrieving and evaluating is $o_r + o_e$. Note that this cost model implies we have some type of index on A so we can reach the examined tuples with constant cost independent of the discarded tuples.

We now present expressions denoting costs across tuples. We first define the following terms for convenience:
- $R_a^+$ denotes the total number of correct tuples that we retrieved from group $a$, while $R_a^-$ denotes the total number of incorrect tuples that we retrieved from group $a$.
- $E_a^+$ denotes the total number of correct tuples that we evaluated from group $a$. $E_a^-$ denotes the total number of incorrect tuples we evaluated from group $a$.

Then, our overall cost, which we will aim to minimize, is:

$$\mathcal{O} = \sum_{a \in A} o_r(R_a^+ + R_a^-) + o_e(E_a^+ + E_a^-) \quad (1)$$

**Metrics and Constraints:** Our output must meet a user-specified *precision* and *recall* constraint. As defined earlier, precision is the fraction of tuples in the output that are correct. Using our notation, $\sum_{a \in A} R_a^+$ is the number of correct tuples that we return, whereas $\sum_{a \in A} R_a^+ + R_a^- - E_a^-$ is the total number of tuples we return (incorrect tuples that are retrieved and evaluated get discarded). Hence,

$$\mathcal{P} = \frac{\sum_{a \in A} R_a^+}{\sum_{a \in A} R_a^+ + R_a^- - E_a^-}$$

Analogously, recall is the fraction of the correct tuples that are present in the output, i.e.,

$$\mathcal{R} = \frac{\sum_{a \in A} R_a^+}{\sum_{a \in A} C_a}$$

The user specifies a precision lower-bound $\alpha \in [0, 1]$, and a recall lower-bound $\beta \in [0, 1]$. Our output must then satisfy:

$$\mathcal{P} \geq \alpha; \ \mathcal{R} \geq \beta$$

A special case of interest is the *browsing* scenario, where 100% precision is required, and thus we have to evaluate every tuple we retrieve, and our objective is to minimize expected cost while satisfying the recall constraint.

**Probabilistic Constraints:** The above bounds are strict, but if we are making discard and evaluate decisions based on imprecise statistics, there is a (hopefully small) chance that the output will violate the bounds. Thus, we also allow the user to define a *satisfaction probability* $\rho$ (which we expect to be very close to 1). The system should then guarantee that with probability $\rho$ the precision and recall constraints are met.

## 3. EXPLOITING CORRELATIONS

In the example of Section 1 we described the two steps of our approach: Step 1 obtains correlations, and Step 2 takes this correlation information as input to process the query efficiently. We consider three forms of such correlations.
- Perfect Information: In this case, Step 2 receives both $W_a$ and $C_a$ (the exact number of wrong and correct tuples, respectively) for all $a$. This case is only included as a baseline, so we can compare the other more realistic cases against it.
- Perfect Selectivities: in this case, we know the probability of each tuple in a group being correct, i.e., $s_a$ for each group $a$. If we take a random tuple from group $a$, then f(ID)=1 for that tuple with probability $s_a$ independent of other tuples. Thus the number of tuples in group $a$ with f(ID)=1 follow a binomial distribution.
- Estimated Selectivities: in this case we do not know the selectivity $s_a$ precisely. That is, we have a random variable $S_a$ that represents the selectivity of group $a$. Selectivity $s_a$ is now the mean of $S_a$ and $v_a$ is the variance of $S_a$.

Exploiting these different types of correlations is the technical focus of this section.

### 3.1 Warm-up: Perfect Information

We begin by considering the *perfect information* case, i.e., we know, in advance, the precise number of correct and incorrect tuples in each group. That is, for each $a \in A$, we know the values of $C_a$, i.e., the number of correct tuples, and $W_a$, the number of incorrect tuples. Thus, in this section, these values are constants.

While these assumptions are not completely realistic (we will rarely have such information available in advance), this section will demonstrate that even when perfect information is available, the problem of selecting which groups to retrieve and evaluate, which groups to retrieve but not evaluate, and which groups to discard is already intractable. Furthermore, the section will act as an introduction to the more complex schemes in subsequent sections.

EXAMPLE 3.1. *To continue our example for section 1, we may know that:*
- *If* `A=1`, `f(ID)` = 1 *for* 900 *out of* 1000 *tuples.*
- *If* `A=2`, `f(ID)` = 1 *for* 500 *out of* 1000 *tuples.*
- *If* `A=3`, `f(ID)` = 1 *for* 100 *out of* 1000 *tuples.*

*If we need 90% precision and recall, then we need to retrieve at least* $1500 \times 0.9 = 1350$ *out of the correct* 1500 *tuples, and at most* 10% *of the retrieved tuples should be incorrect. As an example, if we return all tuples with* `A=1`*, and evaluate all tuples with* `A=2` *and return the ones that have* `f(ID)=1`*, then our solution will contain* $900 + 500 = 1400$ *correct tuples, and* 100 *incorrect tuples, which satisfies both constraints.*

**Query Optimization and Execution:** Since the exact number of correct and incorrect tuples is known, our processing of this query proceeds in two steps:
- **Optimization:** We determine, for each group, whether to (a) retrieve and evaluate, (b) retrieve but not evaluate, or (c) directly discard the tuples in that group. In this section, we focus on *deterministic* schemes; that is, we pick one of (a) (b) or (c) for each group to minimize the cost objective $\mathcal{O}$ while meeting precision and recall constraints. (We will consider probabilistic schemes in subsequent sections.) No query execution happens at this point.
- **Execution:** Then, we execute the query by retrieving all the tuples in the groups for which we either chose to do (a) or (b) in the previous step, then we evaluate all the tuples in the groups for which we chose to do (a).

**Optimization Problem:** Given the values of $C_a$ and $W_a$ for each $a$, the goal of our optimization step is to output boolean decision variables $R_a$ and $E_a$ to satisfy recall and precision constraints while minimizing cost. The boolean variable $R_a$ is set to 1 if the tuples in that group with $A = a$ are to be retrieved during query execution (and 0 if the tuples are to be discarded), and the boolean variable $E_a$ is set to 1 if the tuples in that group with $A = a$ are to be evaluated during query execution (and 0 if they are not). Note that $R_a \geq E_a$ for each $a \in A$ since we may only evaluate the UDF on a tuple if it is retrieved first.

We now rewrite the constraints in the previous section in terms of the new boolean variables $R_a, E_a$. The recall constraint becomes:

$$\sum_{a \in A} C_a R_a \geq \beta \sum_{a \in A} C_a \quad (2)$$

Note that given $C_a$ values, the right hand side of the inequality is a constant $\gamma = \beta \sum_{a \in A} C_a$. The precision constraint becomes:

$$\frac{\sum_{a \in A} C_a R_a}{\sum_{a \in A} C_a R_a + W_a(R_a - E_a)} \geq \alpha$$

$$\Leftrightarrow \sum_{a \in A} ((\frac{1}{\alpha} - 1)C_a - W_a)R_a + W_a E_a \geq 0 \quad (3)$$

Then, the optimization problem can be restated as:

PROBLEM 1 (PERFECT-INFORMATION). *Given* $C_a, W_a, t_a \forall a$, *identify* $R_a, E_a \in \{0, 1\}; R_a \geq E_a$ *to minimize* $\mathcal{O} = \sum_{a \in A} (W_a + C_a)(R_a o_r + E_a o_e)$, *such that Constraints 2 and 3 are satisfied.*

We demonstrate that Problem 1 is NP-hard, which motivates the more relaxed semantics throughout this section.

*Hardness of Special Case.* We can show that Problem 1 is NP-hard even when there is no precision constraint. Since Problem 1 can be cast as an Integer Linear Program (ILP), overall, the problem is NP-Complete.

THEOREM 3.2. *Problem 1 is NP-Complete in $|A|$ (i.e., the number of distinct values of A).*

PROOF. Consider an instance of the minimum knapsack problem. We have a set of objects $S$, with a weight $w_s$ and value $v_s$ for each $s \in S$. In addition, we are given a threshold $V$. Our objective is to find a subset $S' \subset S$ such that $\sum_{s \in S'} v_s \geq V$ while minimising $\sum_{s \in S'} w_s$. This problem is known to be reducible from knapsack problem, and hence NP-Hard. We reduce it to Problem 1. First we can scale up the weights $w_s$ by a constant factor such that $w_s > v_s \forall s \in S$, while keeping the problem effectively unchanged. Then we consider an instance of Problem 1 with $A = S$, and $\forall a \in A : W_a = w_a - v_a, C_a = v_a$. $\alpha$ is set to 0, and $\beta$ to $\frac{V}{\sum_{a \in A} C_a}$. Since $\alpha = 0$, Constraint 3 is trivially satisfied, and the objective becomes ($E_a = 0$ since we don't have any other constraint on $E_a$:

$$\text{Minimize} \sum_{a \in A} C_a R_a, \text{such that} \sum_{a \in A} R_a C_a \geq V$$

. Solving this problem and setting $S' = \{a \in A \mid R_a = 1\}$ gives us a solution for the minimum knapsack problem instance, proving that Problem 1 is NP-Hard. □

Unless otherwise noted, complete proofs of our results (like Theorem 3.1) can be found in our extended technical report [1]. In some selected cases the proofs are included in the appendix of this paper

## 3.2 Perfect Selectivity Information

In the previous section, we demonstrated that the optimization problem to generate deterministic execution schemes meeting precision and recall guarantees exactly, was NP-Hard. In this section we make the following modifications to the problem in the previous section:
- **Selectivities:** We now assume that $W_a, C_a, \forall a \in A$ are not provided to us. Instead, we assume that we know the selectivity $s_a$ of each group $a \in A$, i.e., for a tuple with $A = a$, the probability of it being correct is independently true with probability $s_a$. These selectivities could be learned using historical data. We assume that the selectivity values are *perfect*. (In the next section, we consider selectivity values that may be *imperfect*.)
- **Satisfaction Probability:** Since we are dealing with selectivities which are probabilities, rather than deterministic $W_a$,

$C_a$ values, there are always adversarial inputs on which any query execution strategy would perform poorly. Thus, in this section (and subsequent ones) we assume that the user has given us a probability value $\rho$ such that our precision and recall constraints must be met with probability $\rho$.

In addition, since we are dealing with probabilities for both the quantities mentioned previously, it is more natural to consider probabilistic execution strategies rather than deterministic ones. Accepting fractional solutions to our linear problem, and consequently providing only probabilistic guarantees to error bounds, makes our new optimization problem tractable.

EXAMPLE 3.3. *Continuing our example from earlier, we may know that*
- *If* A=1, f(ID) = 1 *with probability 0.9.*
- *If* A=2, f(ID) = 1 *with probability 0.5.*
- *If* A=3, f(ID) = 1 *with probability 0.1.*

*Each of these probabilities are independent of all other tuples. Say we require a precision of* 90% *and recall of* 90% *with probability at least* 0.9. *We may achieve this by say, (i) returning every tuple with* A=1 *with probability* 1.0, *(ii) evaluating every tuple with* A=2 *with probability* 0.9, *and returning the ones that evaluate to* f(ID)=1 *and (iii) returning the ones with* A=2 *that we did not evaluate.*

We now present the problem formally.

**Query Optimization and Execution:** As before, our processing proceeds in two steps:
- **Optimization:** We determine, for each group, whether to (a) retrieve and evaluate, (b) retrieve but not evaluate, or (c) directly discard the tuples in that group. We do this by generating probabilities $0 \leq E_a, R_a \leq 1, \forall a \in A$. The objective of this step is to choose $R_a$s and $E_a$s such that the results after execution (described next) satisfy the precision and recall constraints with probability at least $\rho$, while minimizing expected cost $\mathcal{O}$. No query execution happens at this point.
- **Execution:** We then execute the query by retrieving each of the tuples in group $a$ with probability $R_a$, independent of all other tuples, and evaluate each of the retrieved tuples from group $a$ with probability $\frac{E_a}{R_a}$, again independently of all other tuples. (Note that these two probabilities when multiplied together give $E_a$, which is precisely the probability we evaluate a tuple from group $a$.) Tuples that are retrieved but not evaluated are returned in the output (that is, we assume those tuples are correct).

**Optimization Problem:** For each $a \in A$, we are given the selectivity $s_a$ of group $a$, i.e., each tuple in group $a$ is correct with probability $s_a$ independent of other tuples.

Our constraints can be expressed as follows: We have the basic constraint $1 \geq R_a \geq E_a \geq 0$. Recall that $R_a^+$ denotes the random variable corresponding to the number of correct tuples retrieved from group $a$ (similarly for $R_a^-, E_a^+, E_a^-$.) The precision constraint, which must be satisfied with probability $\rho$, can be expressed as:

$$\sum_{a \in A} R_a^+ - \alpha \sum_{a \in A}(R_a^+ + R_a^- - E_a^-) \geq 0 \quad (4)$$

The recall constraint, which again must be satisfied with probability $\rho$, can be expressed as:

$$\sum_{a \in A} R_a^+ - \beta \sum_{a \in A} C_a \geq 0 \quad (5)$$

Since our solution is now probabilistic, we try to minimize the expected value of the objective.

$$\mathcal{O} = E\left[\sum_{a \in A} o_r(R_a^+ + R_a^-) + o_e(E_a^+ + E_a^-)\right]$$
$$= \sum_{a \in A}(W_a + C_a)(o_r R_a + o_e E_a) = \sum_{a \in A} t_a(o_r R_a + o_e E_a)$$

PROBLEM 2 (PERFECT-SELECTIVITIES). *Given $s_a, t_a \forall a$, identify $0 \leq E_a \leq R_a \leq 1$ to minimize $\mathcal{O} = \sum_{a \in A} t_a(o_r R_a + o_e E_a)$, such that Constraints 4 and 5 are satisfied with probability greater than $\rho$.*

In the next few subsections, we will (a) describe a LP-based solution for Problem 2 (b) demonstrate that the LP-based solution is asymptotically optimal (c) describe how we may speed up the computation of the LP, by leveraging the special structure of the LP.

We later study the performance of this LP-based solution, referred to as 'Optimal', in the Experiments section.

### 3.2.1 LP-Based Solution

First, notice that the expected value of the LHS of Constraint 4 can be rewritten as follows:

$$G^p = \sum_{a \in A} t_a s_a (1-\alpha) R_a + t_a(1-s_a)\alpha(E_a - R_a) \quad (6)$$

We get the above equation by substituting $E[R_a^+] = s_a t_a R_a$, $E[R_a^-] = (1-s_a)t_a R_a$ (the $E_a$ equations are similar). Similarly, the expected value of the LHS of Constraint 5 can be rewritten as follows:

$$G^r = \sum_{a \in A} t_a s_a R_a - \sum_{a \in A} \beta t_a s_a \quad (7)$$

Our approach will be to ensure that $G^p$ and $G^r$, i.e., the expected values of quantities closely related to precision and recall, are greater than some carefully chosen thresholds $h_\rho^p$ and $h_\rho^r$ respectively, to ensure that the corresponding constraints 4 and 5 are satisfied with probability $\rho$.

We define $h_\rho^p, h_\rho^r$ as follows:

$$h_\rho^p = \sqrt{\frac{\log(1-\rho) \sum_{a \in A} t_a}{2}} \quad (8)$$

$$h_\rho^r = \sqrt{\frac{\log(1-\rho) \sum_{a \in A} t_a(1-\beta)}{2}} \quad (9)$$

The key observation is that the LHS of constraint 4 can be written as a sum of independent random variables, with one random variable per tuple of the table. Thus, Hoeffding's inequality applies [35], and we choose $h_\rho^p$ such that the LHS is within $h_\rho^p$ of its expected value with probability $\geq \rho$ as needed by constraint 4. $h_\rho^r$ is defined similarly for constraint 5.

Then, consider the following linear program:

LINEAR PROG. 3.4 (PERFECT-SELECTIVITIES). *Minimize* $\sum_{a \in A} t_a(R_a o_r + E_a o_e)$ *subject to:*

$$\sum_{a \in A} t_a s_a(1-\alpha) R_a + t_a(1-s_a)\alpha(E_a - R_a) \geq h_\rho^p \quad (10)$$

$$\sum_{a \in A} t_a s_a R_a - \sum_{a \in A} \beta t_a s_a \geq h_\rho^r \quad (11)$$

$$\forall a \in A; 1 \geq R_a \geq E_a \geq 0$$

It can be shown that solving the linear program above gives us $R_a$ and $E_a$ that satisfy the precision and recall constraints with probability at least $\rho$. Our proof uses repeated applications of Hoeffding's inequality, union bounds [35], as well as other standard results in probability. As stated earlier, unless otherwise noted see our extended technical report [1] for proofs.

THEOREM 3.5. *Solving Linear-Prog. 3.3 provides a feasible solution for Problem 2.*

The theorem above only states that the solution will be feasible, but not how close the resulting cost $\mathcal{O}$ is to the optimal cost. Thankfully, as we will show below, our solution is close to optimal.

*Tightness.* We state the tightness bound on our result below, and prove it in the appendix.

THEOREM 3.6. *Let $s_a^{min}$ be the smallest non-zero value of $s_a$. Then, the difference between the cost of the optimal solution to Linear-Prog. 3.3, and the cost of the optimal solution to Problem 2, is at most*

$$(o_e + o_r)\frac{1}{s_a^{min}} \max(h_\rho^r + h_{1-\rho}^r, \frac{h_\rho^p + h_{1-\rho}^p}{1-\alpha})$$

The idea behind this result is as follows: The LHS of constraint 4 in the optimal solution must be greater than $-h_{1-\rho}^p$, because if it is not, then Hoeffding's inequality would imply that the constraint is violated with probability $\geq 1 - \rho$. We can make a similar claim about the recall constraint. Then we can show that by increasing the $R_a$, $E_a$ values of the optimal solution, we can make it satisfy constraints of Linear Program 3.3, while increasing expected cost by at most $(o_e + o_r)\frac{1}{s_a^{min}} \max(h_\rho^r + h_{1-\rho}^r, \frac{h_\rho^p + h_{1-\rho}^p}{1-\alpha})$. The resulting cost must be greater than of equal to the cost of our solution, which was optimal for Linear Program 3.3. This gives us a bound on the cost difference between our solution and the optimal solution to Problem 2.

This cost difference described in the previous theorem is $O(\sqrt{n})$ where $n$ is number of tuples, because the $h^p$, $h^r$ values are $O(\sqrt{n})$. Now consider what happens as $n \to \infty$ for fixed $s_a$'s, $\alpha$, $\beta$. This means we have the same constraints, and groups with the same selectivities, but the group sizes are scaled up. As $n \to \infty$, the ratio of our cost to optimal cost $\to 1$, as indicated by the theorem below:

THEOREM 3.7. *As $n \to \infty$, the ratio of the cost of the optimal solution to Linear-Prog. 3.3, to the cost of the optimal solution to Problem 2 approaches 1.*

Thus, our solution for Problem 2 is asymptotically optimal.

### 3.2.2 BiGreedy Solution for the linear problem

In this section, we describe an efficient way to solve Linear-Prog. 3.3 without even using a linear solver, in $O(|A| \log |A|)$ time:

THEOREM 3.8. *If the following constraints hold:*

$$h_\rho^p < \sum_{a \in A} \max(t_a(s_a - \alpha), 0)$$
$$h_\rho^r < \sum_{a \in A} (1-\beta) t_a s_a$$

*then the solution to Linear-Prog 3.3 can be found in $O(|A| \log |A|)$*
The first constraint simply states that the variance term $h_\rho^p$ is not too large (the constraint is trivially satisfied if the term is 0). It ensures that the precision constraint can be satisfied without evaluating any tuple with selectivity $\geq \alpha$, while the second constraint simply ensures that the problem has a solution.

Our improved algorithm for solving the Linear Program is:

---

Algorithm BIGREEDY-LP
- Initialize all $R_a$ and $E_a$ to 0. Sort the $a$'s based on $s_a$.
- Then start increasing the values of $R_a$ in a greedy fashion. That is, increase $R_a$ for the highest selectivity $a$ until it becomes 1, then increase $R_a$ for the next highest selectivity $a$, and so on. Keep repeating this step until the recall constraint Equation 11 is satisfied.
- After that, keep the $R_a$s fixed, and start increasing the $E_a$s in a greedy fashion, but in reverse order. That is, start by increasing $E_a$ for the $a$ with lowest selectivity and non-zero $R_a$, until it reaches $R_a$, then move to the $a$ with next lowest selectivity, and so on. Keep repeating this step until the precision constraint Equation 10 is satisfied.

---

Thus the approach is to round up the values of the $R_a$ in the order of decreasing $s_a$, and then round up the values of the values of the $E_a$ in the order of increasing $s_a$. The intuition is simple; we want to retrieve all the groups where the selectivity is high, in order to meet our recall constraint. Once we know we're retrieving enough groups to meet our recall constraint, we then ensure that we're evaluating enough groups to meet the precision constraint. Naturally, we'd rather evaluate UDFs for the *most incorrect groups among those we're already retrieving* — i.e., those with the lowest $s_a$ first, so that we can have the maximum impact on precision. The proof that Algorithm BIGREEDY-LP provides the solution to Linear-Prog. 3.3 can be found in the technical report [1].

## 3.3 Estimated Selectivity Information

In the previous section, we developed probabilistic execution strategies when the selectivity of each group is precisely known in advance. In reality, however, we are unlikely to know the precise selectivity of each group, and must instead rely on *estimates of selectivity* (found using sampling or some other method). In this section, we consider the case where we only have an estimate of the selectivity.

As in the previous section, we focus on probabilistic execution strategies that ensure that the precision and recall constraints are met with satisfaction probability $\rho$. Further, as in the previous section, our query processing proceeds in two steps, one, that solves an optimization problem, and second, that actually executes the query using the solution to the optimization problem.

EXAMPLE 3.9. *Continuing our example, suppose we sample some tuples and find that*
- *For* A=1, f(ID) = 1 *for 90 out of 100 sampled tuples.*
- *For* A=2, f(ID) = 1 *for 50 out of 100 sampled tuples.*
- *For* A=3, f(ID) = 1 *for 10 out of 100 sampled tuples.*

*Then for* A=1, *we can guess that the selectivity is close to 0.9, but we cannot say it is exactly equal to 0.9, because it is based on a random sample. Instead, we model the selectivities as a set of independent random variables. In this section, we assume that the random variables are given to us in terms of their mean and standard deviation. In Section 4, we describe how to actually obtain these statistics.*

**Optimization Problem:** Suppose the tuple selectivity for each value $a \in A$ is given by a value $s_a'$ unknown to us. What is known to us instead is an estimate $s_a$ of $s_a'$, which is an instance of a random variable $S_a$. Let $E[S_a] = s_a'$ and $\text{Var}(S_a) \approx v_a$, where $v_a$ is our estimate of the variance of $S_a$. Note that in this section $s_a$ is an estimate of $S_a$, and not the actual selectivity like in the previous section.

We are given $\alpha, \beta, \rho$, and we want to choose probabilities $1 \geq R_a \geq E_a \geq 0, \forall a \in A$, such that when execution strategy retrieves and evaluates tuples probabilistically as described in Section

3.2, the eventual result satisfies the precision and recall constraints with probability at least $\rho$, while minimizing cost:

PROBLEM 3 (ESTIMATED-SELECTIVITIES). *Given $s_a, v_a$, identify $0 \leq E_a \leq R_a \leq 1$ to minimize $\mathcal{O} = \sum_{a \in A} t_a(o_r R_a + o_e E_a)$, such that Constraints 4 and 5 are satisfied with probability greater than $\rho$.*

### 3.3.1 Convex Optimization-Based Solution

Until now, we have not specified the correlations between $S_a$s for different $a$s. We solve the problem for two cases, one where the correlations are unknown (and hence we assume the worst case of maximum correlation between $S_a$s) and the other case where $S_a$s for different $a$'s are independent of each other. Let $e_\rho = \frac{1}{\sqrt{1-\rho}}$. Then our solutions for the two cases are:

CONVEX PROG. 3.10 (UNKNOWN-CORRELATIONS). *Minimize $\sum_{a \in A} t_a(R_a o_r + E_a o_e)$ such that*

$$\sum_{a \in A} (1-\alpha) t_a R_a s_a - t_a \alpha (R_a - E_a)(1 - s_a) \geq X$$

$$\sum_{a \in A} t_a R_a s_a - \beta t_a s_a \geq Y$$

$$e_\rho \sum_{a \in A} \sqrt{v_a} t_a (R_a - \alpha E_a) + 0.5\sqrt{t_a} = X$$

$$e_\rho \sum_{a \in A} \sqrt{v_a} t_a |R_a - \beta| + 0.5\sqrt{t_a} = Y$$

$$R_a, E_a \in \{0, 1\}; R_a \geq E_a$$

Notice that the first two constraints and the objective of the problem above are identical to those in Linear-Prog. 3.3. However, the next two constraints (regarding $X, Y$) are highly non-linear, forcing the problem to be a convex problem. Solving this problem gives a solution which has precision $\geq \alpha$ and recall $\geq \beta$ with probability at least $\rho$ each.

CONVEX PROG. 3.11 (INDEPENDENT GROUPS). *Minimize $\sum_{a \in A} t_a(R_a o_r + E_a o_e)$ such that*

$$\sum_{a \in A} (1-\alpha) t_a R_a s_a - t_a \alpha (R_a - E_a)(1 - s_a) \geq X$$

$$\sum_{a \in A} t_a R_a s_a - \beta t_a s_a \geq Y$$

$$e_\rho \sqrt{\sum_{a \in A} t_a^2 v_a (R_a - \alpha E_a)^2 + 0.25 t_a} = X$$

$$e_\rho \sqrt{\sum_{a \in A} t_a^2 v_a (R_a - \beta)^2 + 0.25 t_a} = Y$$

$$R_a, E_a \in \{0, 1\}; R_a \geq E_a$$

Once again, the first two constraints and the objective of the problem above are identical to those in Linear-Prog. 3.3, while the next two constraints make it highly non-linear.

Solving this problem gives a solution which has precision $\geq \alpha$ and recall $\geq \beta$ with probability at least $\rho$ each. Later in the experiments section, we study the performance of the above solution, referred to as 'Intel-Sample'.

THEOREM 3.12. *Solving Convex-Prog. 3.9 provides a feasible solution for Problem 3 when correlations between $S_a$s are unknown.*

THEOREM 3.13. *Solving Convex-Prog. 3.10 provides a feasible solution for Problem 3 when $S_a$s are independent.*

Let $Q_p$ be the LHS of constraint 4. Using Chebyshev's inequality [35], we argue that to satisfy the constraint $Q_p \geq 0$ with probability $\geq \rho$, it is sufficient to satisfy $E[Q_p] \geq \frac{\text{Dev}(Q_p)}{\sqrt{1-\rho}} = e_\rho \text{Dev}(Q_p)$. The variable $X$ in our convex program is an upper bound on $e_\rho \text{Dev}(Q_p)$, and the LHS of the first convex program constraint is $E[Q_p]$. The other two constraints of the convex program similarly correspond to recall. Complete proofs of the above theorems can be found in the Appendix 10.2.

The above theorems show that our solution is correct, but not how close it is to optimal. However, if we obtain our selectivity estimates by sampling (as described in Section 4), then our method is asymptotically optimal:

THEOREM 3.14. *Let $n$ be the total number of tuples in the table. Suppose we increase $n$, scaling up group sizes while keeping group selectivities fixed. Let the sample size used to obtain selectivity estimates be given by any function that is in $o(n)$. Also let $\mathcal{O}_n^{ind}$ denote the cost of the optimal solution to Convex-Prog. 3.10, $\mathcal{O}_n^{unk}$ the cost of the optimal solution to Convex-Prog. 3.9, and $\mathcal{O}_n^{opt}$ the cost of the optimal solution to Problem 2. Then as $n \to \infty$, $\frac{\mathcal{O}_n^{ind}}{\mathcal{O}_n^{opt}} \to 1$ and $\frac{\mathcal{O}_n^{unk}}{\mathcal{O}_n^{opt}} \to 1$.*

Thus the ratio of cost of our algorithm to the cost of the optimal algorithm tends to 1 as $n$ tends to infinity.

## 4. JOINT ESTIMATION & EXPLOITATION

In the previous section, we developed probabilistic execution strategies when selectivity estimates are already available. But in many cases, selectivity estimates are not usually available beforehand and need to be gathered on-the-fly. In this section, we focus on the problem of jointly estimating and exploiting selectivity information. We will estimate selectivities using sampling (i.e., retrieving and evaluating a small fraction of the tuples).

Our key idea will be to adapt the technique from Section 3.3 to work with sampling-based estimates of selectivities. We first examine how we can map knowledge from samples to the setting from Section 3.3 in Section 4.1. We then describe our extension of the solution from Section 3.3 to this new scenario, assuming samples are given, in Section 4.2. Lastly, we consider the problem of deciding how much to sample from each group in Section 4.3.

### 4.1 Sampling-based Estimates

For the purposes of this subsection, we assume that some number of tuples per group have been sampled, i.e., for each of those tuples, we have retrieved and evaluated them. We now describe how the sampled UDF evaluations lead to selectivity estimates.

For each $a \in A$, we use random variable $S_a$ to represent the distribution of selectivity estimates we may obtain for tuples from group $a$. We assume that the $S_a$s themselves are independent of each other for different $a$s, so evaluating a tuple for $a = a_1 \in A$ tells us nothing about the selectivity $S_{a_2}$, for $a_2 \neq a_1$.

Suppose for each $a \in A$, we have evaluated the UDF on $F_a$ tuples, and found $F_a^+$ of them to satisfy the predicate, and $F_a^-$ that don't (thus $F_a = F_a^+ + F_a^-$). Then at that point, the probability density function of our estimate at any value $x$ will be given by us a Beta distribution [35]: $\text{Beta}(F_a^+ + 1, F_a^- + 1)(x)$. Therefore

$$s_a = E\left[\text{Beta}(F_a^+ + 1, F_a^- + 1)\right] = \frac{F_a^+ + 1}{F_a + 2};$$

$$v_a = \text{Var}(\text{Beta}(F_a^+ + 1, F_a^- + 1)) = \frac{s_a(1 - s_a)}{F_a + 3}$$

where $s_a$ and $v_a$ are defined as per Section 3.3.

### 4.2 Solution given Sampling-based Estimates

As described above, we can use sampling to obtain selectivity estimates. The estimates can then be used to solve the problem in a way similar to Section 3.3.

However, there is a small wrinkle that needs to be dealt with: in addition to giving us selectivity estimates, the sampled tuples will themselves have been evaluated already. So, among these tuples, those that are correct (based on the UDF invocation on that tuple) can be simply returned as part of the query result without re-evaluating them.

Expressing the mean and variance of selectivities in terms of $F_a^+$, $F_a^-$ and $F_a$, and taking into account the tuples that have already been sampled (and hence retrieved and evaluated), we can rephrase the optimization problem from Section 3.3 as below:

CONVEX PROG. 4.1 (SAMPLING SELECTIVITIES). *Minimize* $\sum_{a \in A}(t_a - F_a)(R_a o_r + E_a o_e) + F_a(o_e + o_r)$ *such that*

$$\sum_{a \in A} F_a^+(1-\alpha) + (1-\alpha)(t_a - F_a)R_a s_a -$$
$$(t_a - F_a)\alpha(R_a - E_a)(1 - s_a) \geq X$$

$$\sum_{a \in A} F_a^+ + (t_a - F_a)R_a s_a - \beta(t_a - F_a)s_a \beta F_a^+ \geq Y$$

$$e_\rho \sqrt{\sum_{a \in A}(t_a - F_a)^2 v_a(R_a - \alpha E_a)^2 + 0.25(t_a - F_a)} = X$$

$$e_\rho \sqrt{\sum_{a \in A}(t_a - F_a)^2 v_a(R_a - \beta)^2 + 0.25(t_a - F_a)} = Y$$

$$R_a, E_a \in \{0,1\}; R_a \geq E_a$$

where $e_\rho = \frac{1}{\sqrt{1-\rho}}$. And we have:

THEOREM 4.2. *When $S_a$s are been obtained by sampling $F_a$ tuples from group $a$ for each $a \in A$, of which $F_a^+$ tuples turn out to be correct (based on UDF evaluations for each of the $F_a$ tuples), solving Convex-Prog. 4.1 provides a feasible solution for Problem 3.*

Note that nothing prevents us from going back-and-forth between estimating selectivities and exploiting them: that is, we may continuously update our selectivity estimates as we evaluate more tuples. We can start off with certain selectivity estimates, apply CONVEXPROG 4.1, decide whether to retrieve and/or evaluate for each group, and as we evaluate more tuples per group, we can go back to Section 4.1, derive new estimates for selectivities, and then apply CONVEXPROG 4.1 again: Thus, our algorithms can be used multiple times in an adaptive fashion.

### 4.3 Deciding how much to Sample

We now consider the question of how many tuples $F_a$ to sample from each group $a$. One simple baseline, is to fix a constant $c$ and then sample $c$ tuples from each group. However, since we know the sizes of the different groups, we can use that to significantly reduce overall cost.

We first state a property, and then use that to derive a rule of thumb for how many tuples to sample per group. Let $n$ be the total number of tuples in the table. Then, if the precision and recall thresholds are fixed, it can be shown that a desirable $F_a$ should be

$$O(t_a n^{-\frac{1}{3}})$$

The justification for this statement can be found in the appendix of the technical report. Therefore, our rule of thumb is:

> For a suitably chosen parameter *num*, $F_a$, the number of tuples sampled from group $a$ for each $a \in A$ should be
> $$F_a = num \times t_a n^{-\frac{1}{3}}$$

We can use this rule of thumb in conjunction with the previous sections to decide how much to sample from each group. Naturally, the rule of thumb (as stated) cannot be applied if the value of the parameter *num* is not known to us. As we will see in the experimental section, the rule of thumb is not very sensitive to the value of *num*, and will work even for values of *num* in a fairly large range in a variety of experimental scenarios.

As it turns out (as we will see in the experimental section), the optimal value of *num* is proportional to $\alpha$, the desired precision threshold. In fact, we find that a value $z\alpha$, with $2 \leq z \leq 5$ usually works well. But even if it doesn't, we can guess the optimal value of $z$ using adaptive sampling as follows: We start with a small value of $z\alpha$ for *num*, and keep increasing it. We also keep solving the convex optimization problem (CONVEXPROG. 4.1) for each value of *num* and keep estimating the cost of the solution as we increase *num*. The cost will initially fall as *num* increases, and will later start to rise as *num* crosses its optimal value. When cost starts to rise, we can stop further sampling and proceed to solve the problem using our technique from Section 4.2.

### 4.4 Finding a correlated column

Until now, we have assumed that a correlated column A is known to us. In practical settings, we may not know such a column in advance, and hence we need an efficient way to figure out which column to use. We propose two ways of doing this below.

In both methods, we sample a small fraction, say 1%, of the tuples and obtain the f value for those tuples. We then use these 'labelled tuples' to decide which column to use, in one of two ways:

- We consider every column $x$ of the table, and use the labelled tuples to estimate selectivities for each value of $x$. To avoid overfitting selectivities, we restrict our attention to columns which have $\leq \sqrt{t}$ distinct values if $t$ is the number of sampled tuples. If there are no such columns, we keep increasing $t$ (sampling more tuples) until do get such a column. We then run our algorithm from Section 3.2 for each of the columns, using the estimated selectivities, to get an estimate of the cost. We can then choose the column that gives the minimum cost to be our 'correlated column.' This approach is used to find the correlated column in all of our experiments except the one in Section 6.3.2.
- The 'correlated column' does not need to be a single column from the table. It can be a virtual column created using available data. In this case, we use the labelled data to learn a logistic regression model from the available columns of the table to the UDF f. We then apply the regressor to get a probability score for each unlabelled tuple of the table, and split the tuples into buckets based on their probability scores. The bucket number can then be treated as our correlated virtual column. This approach is used in the experiment in Section 6.3.2.

We observe in the experiments that this procedure adds a negligble runtime cost to our algorithm. The amount of sampling we need later to estimate selectivities is larger than 1%. Also, the 1% labelled tuples can be re-used for both selectivity estimation and as part of the output. These procedures may not select a column with the highest correlation; in this case, this algorithm is still correct but may run more slowly than with the most correlated column.

## 5. EXTENSIONS

We mention some variations of the above problems that can be handled using small extensions to the techniques discussed so far. Further details on the extensions are given in the technical report [1].
**Alternate Objective Functions** There may be scenarios where the user has a fixed cost budget, and wishes to maximize the number of tuples returned while ensuring a lower bound on precision.
**Multiple Predicates** Another variation of the problem is where the query might have multiple chained UDF selects on a table (which is equivalent to a conjunction of multiple UDF predicates). Since precision and recall constraint are specified by the user for the final output, for this variation it may be possible to trade-off accuracy in one predicate for higher accuracy in the other at the same cost. This problem can be handled by introducing one decision variable per mapping of UDF to decisions. For example, there may be one decision variable which is true if and only if we retrieve for the first predicate and retrieve and evaluate the second predicate.
**Single Predicate with Join** We also consider a situation in which the table $T$ being selected on is later joined with another table $T'$. Each tuple of the table $T$ may match with a different number of tuples of $T'$, possibly making it worthwhile for us to evaluate a tuple with low correctness-probability that matches with a large number of tuples from $T'$, over a tuple with higher correctness probability that joins with fewer tuples from $T'$.

## 6. EXPERIMENTS

Our experimental evaluation studies three items: In Experiment 1, we discuss the performance of 'Intel-Sample', our query evaluation algorithm, on a variety of datasets with different selectivities, versus baseline algorithms and algorithms that have perfect knowledge. We show that our algorithm can achieve upto 80% reduction in cost compared to baseline approaches, and is in fact competitive with (unrealistic) approaches that have perfect knowledge of selectivities up front. We then study the relationship between the selectivity of the UDF predicate and the cost savings attained by our algorithm. We also verify that our approaches meet the specified precision and recall constraints. In Experiment 1 evaluation we sample a fixed fraction of tuples. In Experiment 2, we discuss the sensitivity of our approach to the choice of sampling procedure and its parameters. We also demonstrate that our method can be used with more sophisticated sampling models, such as a logistic regressor. In Experiment 3, we discuss the sensitivity of the performance of our approach to the precision, recall, and probability constraints in the query.

### 6.1 Experimental Setup

We use four datasets for our experiments, described below. The selectivities of our datasets are presented in the first column of Table 2. Additional information on the groups of tuples in the datasets is shown in a table in Appendix 10.3.
**Lending Club:** The first dataset, denoted LC, consists of loan data from the peer-to-peer lending website, the Lending Club [2]. The data consists of two csv files, which together contain about 53000 labelled tuples. Each tuple corresponds to a single loan application, and includes attributes like applicant's id, loan amount, term, applicant's grade (a grade assigned by the lending club to reflect borrower quality), employment title, housing status, annual income, purpose of loan. One of the columns is Loan Status, which is the current status of the loan. Values for this column include 'Current', 'Charged Off' (which means that the loan is considered unlikely to be repaid), 'Fully Paid', 'Default' and 'Late'. We assign a value of 'good' to 'Fully Paid' and a value of 'bad' to 'Charged Off', 'Late', and 'Defaulted'.
**Marketing:** The second dataset is related to direct marketing campaigns of a Portuguese banking institution [10, 31]. The campaign was based on phone calls used to get clients to subscribe to a term deposit. The data consists of a csv file containing about 41000 tuples. Each tuple corresponds to a person called during the campaign, and has attributes of the person such as age, job, marital status, and also some social and economic context indicators such as employment variation rate and consumer price index. The last column has value 'y' or 'n', which corresponds to whether the client did or did not subscribe to the term deposit. We consider the value 'y' 'good' and 'n' 'bad'.
**Census:** The third dataset is from the Census Bureau Database [10]. It has about 45000 tuples, each tuple describing one person. The tuple has demographic information such as age, work class, sex and country. The last column says whether or not the person's income is above 50000 per year. We consider $> 50000$ 'good' and $\leq 50000$ 'bad'. This UDF could again be relevant for marketing, since a company might want to preferentially pitch their product to people having higher income.
**Prosper:** The fourth dataset is similar to LC, and consists of loan data from the website Prosper [3]. It consists of about 30000 tuples, each tuple corresponds to an individual loan application. It has columns such as amount borrowed, debt to income ratio, and grade. The Loan Status column has values which we classify as 'good' and 'bad'. 'good' values correspond to loans which were paid back on time, while 'bad' ones are for loans which were either paid late or not repaid at all.

*Protocol.* For each dataset, we have a *designated* attribute for which we specified 'good' and 'bad' values. We assume that the UDF $f$ on each tuple returns the value 'good'/'bad', which is the value of this attribute for that tuple. We consider the query that selects all tuples from the table that have a 'good' value for the designated attribute.

The value of the UDF f for all tuples is known precisely to us for the purposes of evaluation, but assumed to be unknown to any of the query evaluation algorithms initially. We can then simulate 'sampling' by revealing the value of f for any tuple requested by the query evaluation algorithm. Moreover, we can check the precision and recall obtained by any query evaluation algorithm using our knowledge of f. Unless otherwise specified, we set the default value for the query constraints to be $\alpha = 0.8$ (precision), $\beta = 0.8$ (recall), $\rho = 0.8$ (satisfaction probability). We experimented with different costs for evaluation versus retrieval, but found our results were not significantly sensitive to these parameters. Hence we set $o_e = 3$ and $o_r = 1$, which implies that evaluating the UDF is a factor of three more expensive than retrieving the tuple.

### 6.2 Experiment 1 : Performance Comparison

We compare five different query evaluation algorithms. The first two are our algorithms from Section 3 of this paper. The three algorithms after that are baselines:

- Intel-Sample: This is our main algorithm, from Section 3.3. We first use the procedure described in Section 4.4 to decide which correlated column to use. Then we form groups based on that correlated column, sample to get selectivity estimates, and then solve convex problem 3.10 from Section 3.3 to decide what to retrieve and evaluate. For this experiment, we keep the sample size fixed at 5% of the data.
- Optimal: Here we use the technique from Section 3.2. The technique is unrealistic as it requires that selectivities be known

exactly, so we use our knowledge of all values of f to compute the selectivities and provide them as input to the technique. This algorithm has perfect knowledge of selectivities and is a cost lower bound to Intel-Sample.
- **Naive**: In this baseline algorithm, we randomly retrieve $\beta$ fraction of the tuples (where $\beta$ is the recall constraint) and evaluate all of them. This approach only satisfies the recall constraint in expectation, and not with a given probability.
- **Learning (Semi-supervised Learning)**: This is a baseline algorithm that uses semi-supervised learning. Here, we evaluate some tuples, and then use semi-supervised learning to infer the predicate value of the remaining tuples. We then return the tuples that originally evaluated to true as well as those estimated to be true by the semi-supervised classifier. The number of tuples we initially evaluated (labelled training data) is varied until the precision and recall constraints are met. Note that this gives an unfair advantage to the Learning baseline (and also to Multiple), since in practice we do not know how many tuples to evaluate in advance to get the required precision and recall. In spite of this, we will see that our algorithm performs much better than these two baselines.
- **Multiple (Multiple Imputations)**: This is another baseline using semi-supervised learning. However in this case, we use multiple imputations based on class probabilities estimated by the semi-supervised learning algorithm. Once again, we choose the number of tuples to initially evaluate so as to satisfy precision and recall constraints on average across the imputed datasets.

Our method, Intel-Sample, as well as Learning and Multiple, obtain selectivity information via sampling. The cost of sampling tuples to estimate the selectivity is included in the cost of the algorithms.

### 6.2.1 Comparison of Costs

Summary: *Intel-Sample provides up to 80% savings over the naive algorithm, and at the same time, is not much worse than an optimal algorithm that knows all selectivities in advance.*

We estimate the average number of tuples evaluated by Naive, Intel-Sample and Optimal over 50 iterations, and plot them in the bar chart in Figure 1(a). We see that our algorithm (Intel-Sample) performs far better than the Naive baseline in all datasets, with over 80% savings on the LC dataset, while still being not too much worse than Optimal. Since theorem 3.13 implies asymptotically optimality of Intel-Sample, the fact that its cost is not too far from the Optimal algorithm is not surprising.

Summary: *Intel-Sample outperforms two machine learning baselines by up to 60%; the gains reduce as the selectivity reduces.*

Figure 1(b) compares Intel-Sample to the two Machine Learning baselines. For both the baselines (Learning and Multiple), we choose the smallest number of tuples to evaluate that lets us satisfy the precision and recall constraints. Note that these two baselines also satisfy the constraints in expectation and not necessarily with high probability. Yet, Intel-Sample outperforms the baselines by about 60% on LC, about 20% on Prosper and Census, and very slightly on Marketing.

We notice that our savings in general seem to be highest on LC, and lowest on Marketing. This is because of the selectivities of the UDF predicates in the two datasets. Intel-Sample performs better when the selectivity is higher. Table 2 shows the selectivities and savings (relative to Naive and Learning) for each dataset. We note that while savings for very low selectivities are smaller, Intel-Sample still performs marginally better than other algorithms in this worst case, while still giving large savings for moderate and high selectivities.

Intel Sample does not add much in terms of computation cost.

| Dataset | Selectivity | Savings vs. Naive | Savings vs. ML |
|---|---|---|---|
| LC | 0.72 | 81% | 62% |
| Prosper | 0.45 | 43% | 21% |
| Census | 0.24 | 51% | 22% |
| Marketing | 0.11 | 24% | 3% |

*Table 2: Selectivities and Savings for different Datasets*

We measure the computation time taken by our method, including time for finding the best correlated column to use, sampling tuples, learning selectivities, and running the convex optimization. It is less than a 1 second on each of the datasets. This time is insignificant compared to the reduction in UDF evaluations.

In the above experiment, we chose the best correlated column using the scheme in Section 4.4. Sometimes, our scheme may end up choosing a non-optimal column, or we may have a UDF that does not have a strongly correlated column. In order to study the effect of having a weakly correlated column on the performance of our algorithm, we conducted an experiment using different columns (instead of just the "best" column) on the LC dataset. We tried out all columns in the table that had between 1 and 50 distinct values. Column number 8 was the most correlated column. In addition to column 8, we tried using 35 different columns as the correlated column. The number of UDF evaluations required ranges from 9000 for the best column, to 17000 for the worst column. If we had a perfect column which exactly equals the UDF, the cost would have been close to 0. Thus our algorithm seems to perform reasonably well on arbitrary columns, and its cost in our experiments while using column 8 is somewhere in between its cost on the worst column (17000), and its cost on a perfect column ($\approx 0$). Moreover, the cost for even the worst column is much lower than the cost on using the Naive baseline algorithm (45000). In addition, we note two things:

1. If a UDF is very hard to learn, that will reduce the savings of our method, but will also correspondingly reduce the savings of machine learning baselines. Since a machine learning algorithm that guesses the UDF using other columns can be plugged into our algorithm to create a 'virtual' column, our algorithm's savings will only be low in cases that are already hard for other baselines.

2. Even in the worst case with all columns being completely uncorrelated with the UDF, our algorithm still provides a 'correct' set of tuples (satisfying the precision and recall constraints), only without the savings in evaluations. Thus the downside of using our algorithm is bounded by the small computational overhead of running it, while the upside is potentially huge when we do find a column (real or virtual) that is somewhat correlated with the UDF.

### 6.2.2 Satisfaction of Precision and Recall Constraints

While our algorithms avoid several evaluations, they do so at the cost of slightly lower precision and recall. In this experiment, we show that the precision and recall guarantees are met by our algorithm. Our algorithms are supposed to guarantee that the precision and recall constraints will be satisfied with probability at least $\rho$. We now test if this guarantee is being met.

We first fix a value of $\rho$. For that $\rho$ value, we execute the Sampling algorithm 100 times (all the way from sampling, to retrieving and evaluating and returning tuples), and for each execution, we note whether the precision and recall constraints were satisfied or not. Then we compute the accuracy i.e. the fraction of times the constraints were satisfied for this $\rho$, and repeat this procedure for different values of $\rho$. The accuracy is plotted vs. $\rho$, in Figures 2(a) (Precision Accuracy) and 2(b) (Recall Accuracy). The $x = y$ line in the figures is the minimum level of accuracy required to satisfy

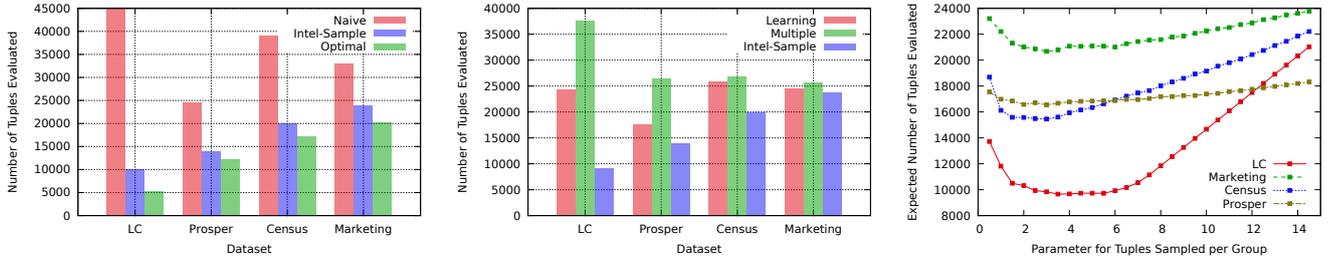

*Figure 1: (a) Number of evaluations for our algorithms compared to baselines (b) Number of evaluations for our algorithms compared with Machine Learning Baselines (c) Expected number of evaluations for different levels of sampling, using logistic regression*

our guarantee. As the figures show, our accuracy is consistently above the line, which means our algorithms do satisfy the precision and recall constraints more than $\rho$ fraction of the time, as desired.

## 6.3 Experiment 2: Robustness of Estimation

As we saw in the previous section, the Intel-Sample algorithm does not involve any unrealistic assumptions like knowing exact selectivities, and still performs far better than the baselines. However, the Intel-Sample algorithm contains free parameters such as the number of tuples to be sampled to obtain from each group to obtain selectivity estimates. In this section, we will attempt to explore how the parameters may be chosen so as to obtain good results.

*Sampling Algorithms.* We compare two different sampling algorithms. Each sampling algorithm tells us how many tuples to sample from each group, before finalizing the selectivity estimates. The output of the sampling algorithm is then used by our algorithm Intel-Sample as input. The sampling algorithms are:

- Constant($c$): This algorithm samples $c$ tuples from each group.
- Two-Third-Power($num$): This algorithm uses our result from Section 4 which says that if a table has $n$ tuples, and group $a$ has $t_a$ tuples, then the optimal number of tuples to sample from group $a$ is proportional to $t_a n^{-\frac{1}{3}}$. Hence this algorithm samples $num \times t_a n^{-\frac{1}{3}}$ tuples from group $a$.

### 6.3.1 Sampling Schemes and Sample Sizes

We now explore the question of how many tuples to sample per group. When we have formed groups but not sampled any tuples, the only information we have is the sizes of the groups.

The schemes of Section 6.3 tell us the proportion in which to sample tuples from each group. Each scheme has an additional free parameter ($c$ or $num$), which we use to determine the exact number of tuples to sample from each group. We study the average cost of evaluating tuples as a function of this parameter, for each scheme.

We first fix the value of the parameter ($c$ or $num$) and use the parameter value to compute the number of tuples to sample from each group. Then we randomly sample and evaluate the corresponding number of tuples from the groups, and form estimates about the selectivity of each group. We then solve the convex optimization problem in Section 3.3 to decide how many tuples to retrieve and evaluate from each group, and compute the expected number of tuples we needed to evaluate to determine cost (the number of tuples retrieved does not vary much for different sampling schemes or sizes). We do this for 100 iterations, and take the average the number of evaluations across iterations. We find average evaluations for several different values of the parameter, and using different correlated columns, and plot the results. The correlated columns chosen and used by our algorithm are Grade for LC and Prosper, Employment Variation Rate for Marketing, and Marital Status for Census. The evaluations vs. parameter $c$ for the **Constant** Sampling scheme is plotted in Figure 3(a), while evaluations vs. parameter $num$ for the **Two-Third-Power** scheme is in Figure 3(b).

We make three observations about Figures 3(a) and 3(b). The first observation is that sampling too little or too much leads to higher cost; too little because our selectivity estimates are bad, and too much because of the cost of sampling itself. There is a region in between that gives low cost. The second observation is that the **Two-Third-Power** scheme generally leads to lower optimal costs than the **Constant** scheme. For instance, the optimal cost for LC is about 10500 for the **Constant** scheme, while it is about 9400 for the **Two-Third-Power** scheme. Similarly, for Census, the optimal cost for scheme **Constant** is about 20000, while for scheme **Two-Third-Power**, the optimal cost is about 19000. The third observation is that a value of $num$ between $2\alpha$ and $5\alpha$ (where $\alpha$ is the precision constraint) leads to near optimal cost for all four datasets, while for the **Constant** scheme, the optimal number of tuples to sample varies quite a bit for different datasets (We discuss further the dependence of optimal $num$ on $\alpha$ in Section 6.4). Thus the **Two-Third-Power** scheme is also robust to small changes in the parameter values, or changes in the predictor or dataset.

While a $num$ parameter value between $2\alpha$ and $5\alpha$ works fine for the current datasets, it may not necessarily be optimal on other datasets. We do not need to know the optimal value of the parameter up front by running our adaptive algorithm in Section 4.

### 6.3.2 Using logistic regression to estimate selectivity

In Section 4.4, we described a technique that uses logistic regression to create a 'virtual column' that can act as a correlated column. Recall that we first randomly sample and evaluate the UDF `f` on 1% of the rows of the table. The resulting set of rows is used as training data to learn a logistic regressor from available columns to the predicate `f`. To avoid overfitting, we only use columns that are either numeric or nominal with $< 50$ different values.

After learning the regressor, we apply it to the table to get a probability score for each tuple (The training tuples are included in the table as well, and the cost of evaluating them before training is taken into account in our cost graphs). Then we split the tuples into 10 buckets (groups) based on the probability score. The bucket ranges are chosen so as to get equal sized buckets. For example, if 10% of the training data tuples got a probability score between 0 and 0.55, and the remaining probability scores were uniformly distributed between 0.55 and 1, then we may create one bucket for tuples with score in $[0, 0.55)$, one for $[0.55, 0.6)$ and so on. Then, we proceed as earlier for these buckets, sampling from them to estimate their selectivities. We do not directly use the selectivity estimates output by the logistic regressor, because we do not have guarantees on the correlation between probability scores of different tuples. (Sampling seems to confirm this lack of correlation; e.g., the fraction of tuples having probability score $\approx 0.5$ which actually satisfy predicate `f` is not close to 0.5). Thus we make buckets and treat them as groups instead, sample from them to estimate selec-

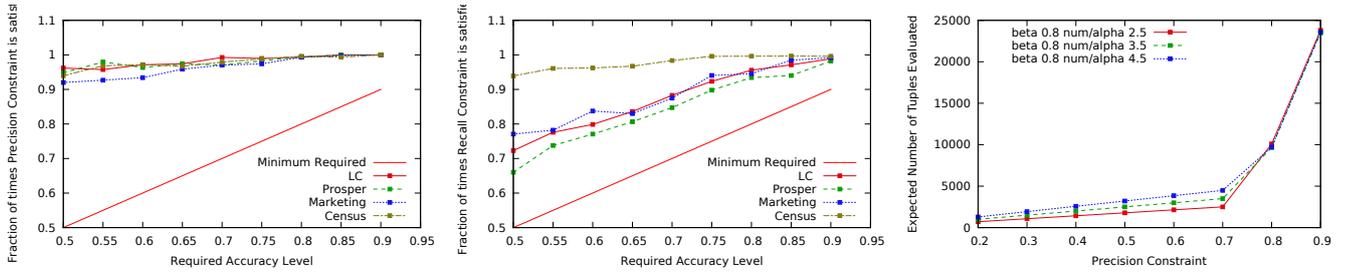

*Figure 2: (a) Fraction of times precision constraint was satisfied, for different values of ρ (b) Fraction of times recall constraint was satisfied, for different values of ρ (c) Expected number of tuples evaluated for different levels of alpha*

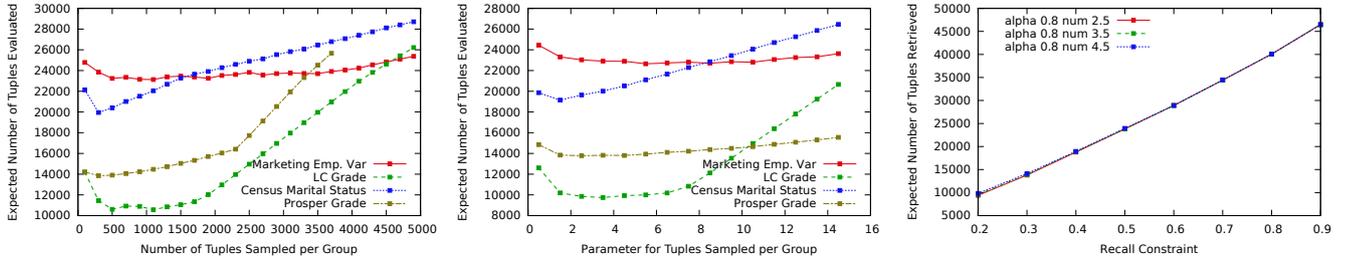

*Figure 3: (a) Expected number of tuples evaluated for different levels if sampling per group for the **Constant** sampling scheme (b) Expected number of tuples evaluated for different levels of sampling per group for the **Two-Third-Power** sampling scheme (c) Expected number of tuples retrieved for different levels of beta*

tivities, and apply the convex optimization from Section 3.3.

In Figure 1(c), we use the **Two-Third-Power** scheme to decide how many tuples to sample per group, and plot the average number of evaluations across 100 iterations, versus the sampling parameter. The number of tuples evaluated as shown in the plot includes the tuples evaluated to generate training data for the logistic regressor. Note that in spite of this, the number of evaluations required by our algorithm is far lower than that required by the Naive algorithm baseline. On LC, we evaluate about 10000 tuples, which is similar to the cost when using a single predictor column. On Census and Marketing, we do noticeably better than when using a single column, while on Prosper, we do slightly worse. But more importantly, using logistic regression to create the correlated column obviates needing to know which column to use.

## 6.4 Experiment 3: Query Parameters

We study the effect of precision and recall constraints ($\alpha$ and $\beta$) on the cost of Intel-Sample and demonstrate that some approximation can result in a large cost saving. To that end, we consider LC with Grade as the fixed predictor, we vary $\alpha$ and $\beta$ from 0.2 to 0.9 each, and we plot average retrieval and evaluation costs over 50 iterations for each combination of $\alpha$ and $\beta$. We keep $\rho$ fixed at 0.8 and use the **Two-Third-Power** sampling scheme for this experiment. The results are similar for other predictors, other $\rho$ values, and other datasets. Figure 2(c) compares the number of evaluations vs $\alpha$ for $\beta = 0.8$, and Figure 3(c) compares the number of retrieves vs $\beta$ for $\alpha = 0.8$. The graph of number of evaluations vs. $\beta$ for $\alpha = 0.8$ is very similar to that in Figure 2(c), while the number of retrieves is nearly independent of $\alpha$.

When $\alpha$ and $\beta$ equal 1 (100%), we must necessarily retrieve and evaluate every tuple, whereas when they're 0, we don't need to retrieve or evaluate any tuple. But when they are both 0.8 we don't need to evaluate 80% of the tuples, but much fewer. The figures show that cost as a function of either $\alpha$ or $\beta$ is a convex function. This tells us why approximating results can be a performance win: a small loss in accuracy can lead to a much larger saving in cost.

As we vary $\alpha$ and $\beta$, we observe that the optimal value of the *num* parameter is approximately proportional to $\alpha$. So we set the value of *num* to $2.5\alpha$, $3.5\alpha$, and so on for the plots. The plots show that $num = 2.5\alpha$ performs slightly better than other values of *num* consistently (for all values of $\alpha$), which reflects our observation that the optimal value of *num* is proportional to $\alpha$. Number of retrieves vs $\alpha$ is not plotted because the two are nearly independent, with number of retrieves depending almost entirely on $\beta$ and $\rho$.

## 7. RELATED WORK

Related work can be divided into a few categories. We describe each one in turn below:

**Sensor Networks:** There has been a lot of work on data processing on sensor networks (see [7] for a survey). We could regard taking additional sensor readings as an expensive operation (which only must be done as needed, say to answer a user query), much like the expensive UDFs we consider in this paper. There has been some work on query processing for sensor networks while minimizing power consumption for data acquisition [30]. This paper uses techniques such as batching of sensor readings, and improved predicate ordering. In addition, there is some work on using correlated attributes to avoid acquiring data [16]. This work focuses on queries with a conjunction of multiple expensive predicates, and uses correlations to determine the optimal predicate order to minimize the expected number of predicate evaluations. The above works still attempt to solve the query exactly, rather than obtain an approximation at a reduced cost.

**Approximate Query Processing:** Over the past decade, there has been a lot of work on approximate query processing; see [13, 19, 21, 27]. Garofalakis et al. [18] provides a good survey of the area. There are multiple systems that support approximate query processing, including BlinkDB [6] and Aqua [5], allowing users to trade-off accuracy for performance. Often, these schemes achieved the desired level of accuracy by running the query on a sample of the database. There has also been work on deciding how to choose appropriate samples for query processing [4, 9]. All of these systems and algorithmic papers focus only on approximating numerical aggregate quantities, having operators such as SUM, AVG, VAR, as well as frequency moments [8]. On the other hand, our paper focuses on approximating selection (i.e., set-valued) queries. Ioannidis et al. [26] use histograms to approximate set-valued queries, but their notion of approximation involves approximating values within tuples not with repsect to precision and recall constraints.

**Query Processing with UDFs:** UDFs occur often in database queries. The work on optimizing queries with UDFs has focused mainly on identifying optimal placement of UDFs in query plans [15, 24], and not on using correlations to avoid UDF evaluations in the first place. Recently, crowd-powered database systems, such as CrowdDB [17] and Deco [34] have been treating calls to human workers for data processing and data gathering on-the-fly, as expensive UDFs that should be used sparingly. Our optimization techniques could easily be applied to these systems.

**Correlation-aware Query Processing:** There has been some recent work on finding correlations between attributes of a relation [12, 25, 28] — these papers could be used to identify the correlated attributes to bootstrap our techniques for estimating and exploiting correlations. Correlations have also been used for traditional query optimization, allowing for more accurate estimates of intermediate result sizes, and therefore of query-plan costs. Recent work has also looked at appropriately choosing which indexes and materialized views to maintain taking correlations into account [29]. While these papers mainly focus on correlations between attributes within a relation (and therefore there is no cost to acquiring data since it is already provided), our paper exploits correlations between attributes in a relation and an external UDF.

## 8. CONCLUSION

We introduced techniques for efficiently approximating select queries with expensive UDF predicates. We found that even though the basic problem for even the simplest scenario is NP-Hard, there are efficient techniques that allow us to achieve near-optimality. Our techniques apply to a variety of different scenarios, based on what information is available to us, and allow us to reduce cost while satisfying precision and recall constraints specified by the user. As UDFs become increasingly ubiquitous and datasets get larger, the cost savings can become quite significant. Moreover, our methods integrate well with machine learning techniques such as logistic regression, allowing us to leverage these techniques to reduce cost even further. Our techniques can also be generalized to other classes of queries, such as selects followed by joins, or multiple selection conditions. Finally, our experiments on real data show that we can obtain cost savings upto $80\%$, which are quite large relative to the small loss in accuracy. As future work, we plan to explore generalizing our techniques to complete select-from-where queries, as well as queries involving nesting and aggregation.

## 10. APPENDIX

### 10.1 Correctness Proof for our solution to Perfect Selectivities Case

$R_a^+$, $R_a^-$, $E_a^+$, $E_a^-$, $C_a$ and $W_a$ are all binomial random variables, since they're a sum of Bernoulli random variables. Moreover, since each tuple with $A = a$ is retrieved randomly with probability $R_a$ independently of other tuples, and correct with probability $s_a$ independently of other tuples, we must have

$$E\left[R_a^+\right] = t_a s_a R_a$$

$$Var(R_a^+) = t_a s_a R_a (1 - s_a R_a)$$

and

$$E\left[R_a^-\right] = t_a (1 - s_a) R_a$$

$$Var(R_a^-) = t_a (1 - s_a) R_a (1 - (1 - s_a) R_a)$$

Similarly,

$$E\left[E_a^+\right] = t_a s_a E_a$$

$$Var(E_a^+) = t_a s_a E_a (1 - s_a E_a)$$

and
$$E[E_a^-] = t_a(1-s_a)E_a$$
$$Var(E_a^-) = t_a(1-s_a)E_a(1-(1-s_a)E_a)$$
.

Thus for the precision, we have:
$$E\left[\sum_{a\in A} R_a^+ - \alpha \sum_{a\in A}(R_a^+ + R_a^- - E_a^-)\right]$$
$$= \sum_{a\in A} t_a s_a R_a(1-\alpha) + \alpha t_a(1-s_a)(E_a - R_a)$$

For recall, we have,
$$E\left[\sum_{a\in A} R_a^+ - \beta \sum_{a\in A} C_a\right] = \sum_{a\in A} t_a s_a R_a - \beta t_a s_a$$

To convert the probabilistic satisfaction constraints for precision/recall, into an absolute constraint in terms of the $R_a$s and $E_a$s, we need the general form of Hoeffding's inequality:

**Hoeffding's inequality:** Let $X_1$, $X_2$, ... $X_n$ be independent random variables, which are almost surely bounded i.e.
$$Pr(a_i \leq X_i \leq b_i) = 1$$
Let there sum be
$$S = \sum_{i=1}^n X_i$$
Then:
$$Pr(S - E[S] \leq -t) \leq e^{-\frac{2t^2}{\sum_{i=1}^n (b_i - a_i)^2}}$$
$$Pr(S - E[S] \geq t) \leq e^{-\frac{2t^2}{\sum_{i=1}^n (b_i - a_i)^2}}$$
□

For a tuple $t$ in the table that satisfies the predicate, consider the variable $Ip_t$ which takes value:

1. $1-\alpha$ if the tuple is correct and retrieved
2. $-\alpha$ if the tuple is incorrect and retrieved but not evaluated
3. 0 otherwise

Then the lhs of the precision constraint, is the sum of $Ip_t$ over all $t$ in the table.

Similarly, for every tuple $t$ let $Ir_t$ be an indicator variable which takes value:

1. $1-\beta$ if the tuple is correct and retrieved
2. 0 otherwise

Then the lhs of the recall constraint is the sum of $Ir_t$ over all $t$.

All the $Ip$ variables are independent of each other, and all the $Ir$ variables are independent of each other. Moreover, the variables are all bounded in the following ranges:

1. $Ip$ in $[-\alpha, 1-\alpha]$
2. $Ir$ in $[0, 1-\beta]$

Now we can apply Hoeffding's inequality to get a concentration bound on the lhs of the precision and recall constraints. Specifically,
$$Pr(|(\sum_{a\in A} R_a^+ - \alpha \sum_{a\in A}(R_a^+ + R_a^- - E_a^-)) -$$
$$(\sum_{a\in A} t_a s_a(1-\alpha)R_a + t_a(1-s_a)\alpha(E_a - R_a))|$$
$$\geq t) \leq e^{-\frac{2t^2}{\sum_{a\in A} t_a}}$$

and
$$Pr(|(\sum_{a\in A} R_a^+ - \beta C_a) - (\sum_{a\in A} t_a s_a R_a - \beta t_a s_a)| \geq t) \leq e^{-\frac{2t^2}{\sum_{a\in A} t_a(1-\beta)}}$$

We have
$$h_\rho^p = \sqrt{\frac{log(1-\rho)\sum_{a\in A} t_a}{2}}$$
and
$$h_\rho^r = \sqrt{\frac{log(1-\rho)\sum_{a\in A} t_a(1-\beta)}{2}}$$

Then with probability $\geq \rho$ the lhs of the precision constraint will be within $h_\rho^p$ of it's expectation, and the lhs of the recall constraint will be within $h_\rho^r$ of its expectation.

Our $R_a$s and $E_a$s satisfy
$$\sum_{a\in A} t_a s_a(1-\alpha)R_a + t_a(1-s_a)\alpha(E_a - R_a) \geq h_\rho^p$$

So with probability $\geq \rho$, the value of $\sum_{a\in A} R_a^+ - \alpha\sum_{a\in A}(R_a^+ + R_a^- - E_a^-)$ will be greater than it's expectation minus $h_\rho^p$, that is, $\sum_{a\in A} t_a s_a(1-\alpha)R_a + t_a(1-s_a)\alpha(E_a - R_a) - h_\rho^p$ which is greater than 0.

Our $R_a$s and $E_a$s also satisfy
$$\sum_{a\in A} t_a s_a R_a - \sum_{a\in A} \beta t_a s_a \geq h_\rho^r$$

So with probability $\geq \rho$, the value of $\sum_{a\in A} R_a^+ - \beta C_a$ will be greater than it's expectation minus $h_\rho^r$, i.e. $(\sum_{a\in A} t_a s_a R_a - \beta t_a s_a) - h_\rho^r$ which is greater than 0.

Thus both precision and recall constraints are satisfied with probability $\rho$.

## 10.2 Proof of Tightness Bound for solution to Perfect Selectivities Case

Suppose an optimal solution (where precision and recall constraints are satisfied with probability $\rho$ and expected cost is minimised) is given by $R_a^o$, $E_a^o$ for each $a \in A$. The expected cost for this solution is $\sum_{a\in A} t_a(R_a^o o_r + E_a^o o_e)$. Consider the expected values of lhs of their precision and recall constraints. For precision, it is given by:
$$P^o = \sum_{a\in A} t_a s_a(1-\alpha)R_a^o + t_a(1-s_a)\alpha(E_a^o - R_a^o)$$

and for recall, it is given by:
$$R^o = \sum_{a\in A} t_a s_a R_a^o - \beta t_a s_a$$

Let the solution produced by our algorithm by given by $R_a'$, $E_a'$ for each $a \in A$. And let:
$$P' = \sum_{a\in A} t_a s_a(1-\alpha)R_a' + t_a(1-s_a)\alpha(E_a' - R_a')$$

and for recall, it is given by:

$$R' = \sum_{a \in A} t_a s_a R'_a - \beta t_a s_a$$

Hence $P' \geq h_\rho^p$, $R' \geq h_\rho^r$

Moreover, we must have $P^o \geq -h_{1-\rho}^p$, since if the expected value of the lhs of precision constraint for the optimal solution is less than $-h_{1-\rho}^p$, then by the chernoff bound, the lhs of the precision constraint will be $\leq 0$ with probability $\geq 1 - \rho$, and hence the precision constraint will be satisfied with probability $\leq \rho$. For a similar reason, we must have $R^o \geq -h_{1-\rho}^r$.

We will now show two tightness bounds, one in terms of the cost, and one in terms of the precision and recall constraints.

First we will find a bound in terms of expected cost $\sum_{a \in A}(t_a s_a + t_a(1-s_a))(R_a o_r + E_a o_e)$. We know that

$$\sum_{a \in A} R_a^o t_a s_a - \sum_{a \in A} \beta t_a s_a \geq -h_{1-\rho}^r$$

and

$$\sum_{a \in A}(1-\alpha)t_a s_a R_a^o + t_a(1-s_a)\alpha(E_a^o - R_a^o) \geq -h_{1-\rho}^p$$

. We will increase some values of $R_a$, and increase the corresponding values of $E_a$ by the same amount, so as to get the resulting $R_a$ and $E_a$ values to satisfy, $\sum_{a \in A} R_a^{o2} t_a s_a \geq h_\rho^r + \sum_{a \in A} \beta t_a s_a$ and $\sum_{a \in A}(1-\alpha)t_a s_a R_a^o + t_a(1-s_a)\alpha(E_a^o - R_a^o) geq h_\rho^p$. We choose which $R_a$s value to raise, in a greedy manner, as follows: We choose the $a$ with highest value of selectivity ($\frac{t_a s_a}{t_a s_a + t_a(1-s_a)}$) for which $R_a < 1$, and keep increasing that $R_a$ (and $E_a$) till $R_a$ becomes 1 (or both the constraints are satisfied), then move on to the next best $a$, and so on. When a $R_a$ and $E_a$ are increased by $d$, the value of $\sum_{a \in A} R_a t_a s_a$ goes up be $dt_a s_a$, the value of $\sum_{a \in A}(1-\alpha)t_a s_a R_a + t_a(1-s_a)\alpha(E_a - R_a)$ goes up by $d(1-\alpha)t_a s_a$ while cost goes up by $(o_r + o_e)d(t_a s_a + t_a(1-s_a))$. If we keep raising the values till both constraints $\sum_{a \in A} R_a^{o2} t_a s_a \geq h_\rho^r + \sum_{a \in A} \beta t_a s_a$ and $\sum_{a \in A}(1-\alpha)t_a s_a R_a^o + t_a(1-s_a)\alpha(E_a^o - R_a^o) geq h_\rho^p$ are satisfied, then the cost increase will be at most

$$(o_e + o_r)\frac{1}{s_a} Max(h_\rho^r + h_{1-\rho}^r, \frac{h_\rho^p + h_{1-\rho}^p}{1-\alpha})$$

where $s_a$ is the smallest value of selectivity from among the $a$s whose $R_a, E_a$ we raised. Now since the resulting $R_a$ and $E_a$ values satisfy these constraints, the total cost $\sum_{a \in A}(t_a s_a + t_a(1-s_a))(R_a o_r + E_a o_e)$ has to be more than the cost of our solution $(R'_a, E'_a)$ (which was optimal for the linear problem with the given constraints). Thus the difference between the cost of our solution, and the cost of the optimal solution to the original problem, is at most

$$(o_e + o_r)\frac{1}{s_a} Max(h_\rho^r + h_{1-\rho}^r, \frac{h_\rho^p + h_{1-\rho}^p}{1-\alpha})$$

. This is our tightness bound on cost. If table size is $n$, and the minimum $s_a$ is constant as $n$ grows, the cost difference is $O(\sqrt{n})$ because the $h^p, h^r$ values are $O(\sqrt{n})$. Since the cost of the solution is expected to be linear in $n$, an extra cost of $O(\sqrt{n})$ should be a small fraction of the actual cost.

For the precision/recall tightness bound, let Let $\beta^\star = \beta + \frac{h_\rho^r + h_{1-\rho}^r}{\sum_{a \in A} t_a s_a}$. And let $\alpha^\star = \alpha + \frac{h_\rho^p + h_{1-\rho}^p}{\sum_{a \in A} \beta^\star t_a s_a}$ And let the optimal solution for precision $\alpha^\star$ (with probability $\rho$) and recall $\beta^\star$ (with probability $\rho$) be $R_a^\star, E_a^\star$ for each $a \in A$. Then that optimal solution must satisfy:

$$\sum_{a \in A}(1-\alpha^\star)t_a s_a R_a + \alpha^\star t_a(1-s_a)(E_a - R_a) \geq -h_{1-\rho}^p$$

and

$$\sum_{a \in A} R_a t_a s_a \geq -h_{1-\rho}^r + \sum_{a \in A} \beta^\star t_a s_a$$

which implies (because of our choice of $\alpha^\star$ and $\beta^\star$)

$$\sum_{a \in A}(1-\alpha)t_a s_a R_a + \alpha t_a(1-s_a)(E_a - R_a) \geq h_\rho^p$$

and

$$\sum_{a \in A} R_a t_a s_a \geq h_\rho^r + \sum_{a \in A} \beta t_a s_a$$

Hence the optimal solution for $\alpha^\star, \beta^\star$ satisfies the constraints of the linear problem we used to find $R'_a, E'_a$. Since we had the optimal solution to our linear problem, it's cost is necessarily less than the cost of $R_a^\star, E_a^\star$. Thus our solution has cost less than the optimal cost for $\alpha^\star$ and $\beta^{star}$. Note that for table size $n$, the $h^p, h^r$ terms grow as $O(\sqrt{n})$, while the $\sum_{a \in A} t_a s_a$ would probably be linear in $n$. Thus the difference between $\beta$ and $\beta^\star$, and the difference between $\alpha$ and $\alpha^\star$ are in $O(\frac{1}{\sqrt{n}})$.

## 10.3 Proof for Greedy solution to the linear problem

For purposes of this proof, let's use $c_{a_i}$ to denote $t_{a_i} s_{a_i}$ and $w_{a_i}$ to denote $t_{a_i}(1-s_{a_i})$ for all $i$.

**Lemma:** Suppose we have $a_1$ and $a_2$ such that $s_{a_1} > s_{a_2}$. Then in an optimal solution, one of the following must be true:

1. $E_{a_2} = R_{a_2}$
2. $E_{a_1} = 0$

**Proof:** Suppose $E_{a_2} < R_{a_2}$ and $E_{a_1} > 0$. Suppose we replace $E_{a_2}$ by $E_{a_2} + \epsilon$, and $E_{a_1}$ by $E_{a_1} - \epsilon\frac{w_{a_2}}{w_{a_1}}$ for an epsilon small enough so as to not violate the $0 \leq E_a \leq R_a$ bound for either $a_1$ or $a_2$. Then the recall constraint and rhs of precision constraint aren't affected at all, while lhs of precision constraint is increased by $\epsilon\alpha(w_{a_2} - \delta w_{a_1}) = 0$. Thus both the constraints continue to be satisfied as before. On the other hand, cost increases by $o_e\epsilon(w_{a_2}+c_{a_2}-\frac{(w_{a_2})(w_{a_1}+c_{a_1})}{w_{a_1}}) = o_e\epsilon(c_{a_2}-\frac{c_{a_1}w_{a_2}}{w_{a_1}}) < 0$. Thus this transformation keeps the constraints satisfied and reduces cost, contradicting the optimality of the solution. Hence, at least one the the conditions

1. $E_{a_2} = R_{a_2}$
2. $E_{a_1} = 0$

must be satisfied.

This lemma shows why once the $R$s are fixed, the $E$s must be filled in increasing selectivity order. According to the lemma, $E$ for at most one selectivity value can be strictly between 0 and $R$, while $E$s must be zero for all higher selectivities, and $R$ for all lower selectivities.

**Lemma:** Suppose we have $a_1$ and $a_2$ such that $s_{a_1} > s_{a_2}$. Then in an optimal solution, one of the following must be true:

1. $R_{a_1} = 1$
2. $R_{a_2} = 0$

**Proof:** Suppose $R_{a_1} < 1$ and $R_{a_2} > 0$. We consider two cases:

1. **Case 1**: $E_{a_2} = R_{a_2}$. Reduce $R_{a_2}, E_{a_2}$ by $\epsilon$, and raise $R_{a_1}$, $E_{a_1}$ by $\epsilon\frac{c_{a_2}}{c_{a_1}}$. This keeps the rhs and lhs values of both the constraints unchanged, but reduces cost by $(w_{a_2} + c_{a_2})(o_r + o_e)\epsilon - (w_{a_1} + c_{a_1})(o_r + o_e)\epsilon \frac{c_{a_2}}{c_{a_1}} = \epsilon(o_r + o_e)(w_{a_2} - w_{a_1} \frac{c_{a_2}}{c_{a_1}}) > 0$. This contradicts the optimality of the solution, proving the lemma in case 1.

2. **Case 2**: $E_{a_2} < R_{a_2}$. Let $l = max(\frac{c_{a_2}}{c_{a_1}}, \frac{(1-\alpha)c_{a_2} - \alpha w_{a_2}}{(1-\alpha)c_{a_1} - \alpha w_{a_1}})$ Reduce $R_{a_2}$ by $\epsilon$, and raise $R_{a_1}$ by $l\epsilon$. The rhs of both constraints remain the same, while the lhs either remain the same or increase. Hence the constraints continue to be satisfied. The cost decreases by $\epsilon o_r(w_{a_2} + c_{a_2} - lw_{a_1} - lc_{a_1})$. If $l = \frac{c_{a_2}}{c_{a_1}}$, then cost decreases as shown in case 1. If $l = \frac{(1-\alpha)c_{a_2} - \alpha w_{a_2}}{(1-\alpha)c_{a_1} - \alpha w_{a_1}}$, then cost decreases by

$$\epsilon o_r \frac{1}{(1-\alpha)c_{a_1} - \alpha w_{a_1}}$$
$$((w_{a_2} + c_{a_2})((1-\alpha)c_{a_1} - \alpha w_{a_1}) - (w_{a_1} + c_{a_1})((1-\alpha)c_{a_2} - \alpha w_{a_2}))$$
$$= \epsilon o_r \frac{c_{a_1} w_{a_2} - c_{a_2} w_{a_1}}{(1-\alpha)c_{a_1} - \alpha w_{a_1}}$$
$$> 0$$

This contradicts the optimality of the solution, thus proving the lemma.

The second lemma means that there is at most one selectivity for which $0 < R < 1$, and $R$ is 0 for all lower selectivities and 1 for all higher selectivities.

**Lemma:** In an optimal solution, either the precision constraint must be satisfied tightly (with equality) or all the $E$s must be 0. In addition, the recall constraint must be satisfied tightly.

**Proof:** Suppose the precision constraint is not tight in our solution, and at least one $E$ is $> 0$. Then we can decrease that $E$ slightly, while still satisfying the precision and recall constraints, and reducing cost. This proves the first part of the lemma.

We next prove that the recall constraint must also be tight. Let $a_r$ be the lowest selectivity $a$ with $R > 0$. Suppose the recall constraint isn't tight, suppose all $E$s are 0. If the precision-constraint-coefficient $(1-\alpha)c_{a_r} - \alpha w_{a_r} < 0$, then decreasing $R_{a_r}$ slightly will continue to satisfy the precision and recall constraints and reduce cost, which is not possible. So recall constraint must be tight if all $E$s are 0.

That leaves the case where not all $E$s are zero (so precision constraint is tight) and recall constraint isn't tight. Let $a_r$ be the lowest selectivity $a$ with $R > 0$ and let $a_e$ be the highest selectivity $a$ with $E < R$. Because of condition $X < \sum_{a \in A} max((1-\alpha)t_a s_a - \alpha t_a(1-s_a), 0)$, the selectivity of $a_e$ must be $<= \alpha$. For a small $\epsilon$, decreasing $R_{a_r}$ and $E_{a_r}$ by $\epsilon$ and increasing $E_{a_e}$ by $\epsilon \frac{(1-\alpha)c_{a_r}}{\alpha w_{a_e}}$ will continue to satisfy the precision and recall constraints. Cost will reduce by $\epsilon o_r(c_{a_r} + w_{a_r}) + \epsilon o_e(w_{a_r} + c_{a_r} - \frac{(w_{a_e}+c_{a_e})(1-\alpha)c_{a_r}}{\alpha w_{a_e}})$. Clearly, $\epsilon o_r(c_{a_r}+w_{a_r}) > 0$. And $(\alpha w_{a_e})(w_{a_r}+c_{a_r}) - (1-\alpha)c_{a_r}(w_{a_e} + c_{a_e})$ $\geq (w_{a_e} + c_{a_e})(w_{a_r} + c_{a_r})(\alpha(1-\alpha) - (1-\alpha)\alpha) = 0$. Thus the transformation keeps constraints satisfied and reduces cost, which contradicts the optimality of the solution, proving the lemma.

Thus the recall constraint must be tight. This, along with the lemma on greedy assignment of $R$s, proves that increasing $R$s in decreasing order of selectivity, until the recall constraint is satisfied, is optimal. After that, the greedy assignment of $E$s lemma shows that increasing $E$s in order of increasing selectivity gives the optimal solution.

## 10.4 Correctness Proof for our solution to the Estimated Selectivities Case

As before, we have for each $a \in A$:

1. $R_a^+$ is the number of correct tuples ($f(B) = 1$) that we end up retrieving for $A = a$.
2. $R_a^-$ is the number of incorrect tuples that we end up retrieving for $A = a$.
3. $E_a^+$ is the number of correct tuples we end up evaluating (and hence accepting) for $A = a$.
4. $E_a^-$ is the number of incorrect tuples we end up evaluating (and hence rejecting) for $A = a$.

In addition, we have

1. $C_a$ is the number of tuples satisfying the predicate for $A = a$.
2. $W_a$ is the number of tuples not satisfying the predicate for $A = a$.

Then we must satisfy

$$\sum_{a \in A} R_a^+ - \alpha \sum_{a \in A}(R_a^+ + R_a^- - E_a^-) \geq 0$$

must be satisfied with probability $\rho$, and

$$\sum_{a \in A} R_a^+ - \beta C_a \geq 0$$

must be satisfied with probability $\rho$.

Let

$$P = \sum_{a \in A} R_a^+ - \alpha \sum_{a \in A}(R_a^+ + R_a^- - E_a^-)$$

and let

$$R = \sum_{a \in A} R_a^+ - \beta C_a$$

To satisfy the constraints $P \geq 0$ and $R \geq 0$ with probability $\rho$ each, we want to satisfy

$$E(P) \geq e_\rho \text{Dev}(P)$$

and

$$E(R) \geq e_\rho \text{Dev}(R)$$

for an appropriate $e_\rho$ determined by Chebyshev's inequality. Thus we want to find the expectation and variances of $P$ and $R$ in terms of $R_a$, $E_a$, and fixed quantities.

For a random variable $U$ (from among $R_a^+$, $R_a^-$, $E_a^+$, $E_a^-$, $C_a$, $W_a$), and a tuple number $i$, let $I_{U,i}$ denote the indicator variable that is 1 if the $i^{th}$ tuple adds one to $U$, and 0 otherwise. For example, $I_{R_a^+,2}$ is 1 if tuple number 2 for $A = a$ is retrieved and satisfies the predicate, and 0 otherwise. Thus $U = \sum_{i=1}^{t_a} I_{U,i}$. The $I$'s are all Bernoulli random variables. Both $P$ and $R$ can be expressed as weighted sums of these Bernoulli random variables. Thus to find expectation and variance of $P$ and $R$, we first try to find the expectation, variance, and covariances of the $I$ variables. Each tuple with $A = a$ is retrieved with probability $R_a$, and evaluated with probability $E_a$. Moreover, it satisfies the predicate with probability $S_a$. Thus this tuple contributes 1 to $R_a^+$ with probability $R_a S_a$. So for each $i$, $I_{R_a^+,i}$ is a Bernoulli random variable with

$$E\left[I_{R_a^+,i}\right] = R_a S_a$$

. Similarly,

$$E\left[I_{R_a^- - E_a^-,i}\right] = (R_a - E_a)(1 - S_a)$$

$$E\left[I_{C_a,i}\right] = S_a$$

For all $U, i$,
$$Var(I_{U,i}) = E\left[I_{U,i}\right](1 - E\left[I_{U,i}\right])$$

For each $U, i$, $\text{Var}(I_{U,i} = E\left[I_{U,i}\right](1 - E\left[I_{U,i}\right])$

To find covariances, we note that
$$E\left[S_a^2\right] = E\left[(1-S_a)^2\right] = s_a^2 + v_a$$

and
$$E[(1-S_a)S_a] = s_a(1-s_a) - v_a$$

Thus for $i \neq j$
$$\begin{aligned}\text{Cov}(I_{R_a^+,i}, I_{R_a^+,j}) &= E\left[I_{R_a^+,i} I_{R_a^+,j}\right] - E\left[I_{R_a^+,i}\right] E\left[I_{R_a^+,j}\right] \\ &= E\left[R_a S_a R_a S_a\right] - R_a s_a R_a s_a \\ &= R_a^2(v_a + s_a^2) - R_a^2 s_a^2 \\ &= R_a^2 v_a\end{aligned}$$

Similarly,
$$\text{Cov}(I_{R_a^- - E_a^-, i}, I_{R_a^- - E_a^-, j}) = (R_a - E_a)^2 v_a$$

and
$$\text{Cov}(I_{C_a,i}, I_{C_a,j}) = v_a$$

Moreover,
$$\begin{aligned}\text{Cov}(I_{R_a^+,i}, I_{R_a^- - E_a^-, j}) &= E\left[I_{R_a^+,i} I_{R_a^- - E_a^-, j}\right] - E\left[I_{R_a^+,i}\right] E\left[I_{R_a^- - E_a^-, j}\right] \\ &= E\left[R_a S_a (R_a - E_a)(1 - S_a)\right] \\ &\quad - R_a s_a (R_a - E_a)(1 - s_a) \\ &= R_a(R_a - E_a)(s_a(1-s_a) - v_a) \\ &\quad - R_a(R_a - E_a)s_a(1 - s_a) \\ &= -R_a(R_a - E_a)v_a\end{aligned}$$

and similarly,
$$\text{Cov}(I_{R_a^+,i}, I_{C_a,j}) = R_a v_a$$

Finally,
$$\begin{aligned}\text{Cov}(I_{R_a^+,i}, I_{R_a^- - E_a^-, i}) &= E\left[I_{R_a^+,i} I_{R_a^- - E_a^-, i}\right] - E\left[I_{R_a^+,i}\right] E\left[I_{R_a^- - E_a^-, i}\right] \\ &= E[0] - R_a s_a (R_a - E_a)(1 - s_a) \\ &= -R_a(R_a - E_a)s_a(1-s_a)\end{aligned}$$

and
$$\begin{aligned}\text{Cov}(I_{R_a^+,i}, I_{C_a,i}) &= E\left[I_{R_a^+,i} I_{C_a,i}\right] - E\left[I_{R_a^+,i}\right] E\left[I_{C_a,i}\right] \\ &= E\left[R_a S_a\right] - R_a s_a s_a \\ &= R_a s_a (1 - s_a)\end{aligned}$$

Now that we know the relevant expectations, variances and covariances, we can compute the expectation and variance of $P$ and $R$. We have
$$P = \sum_{a \in A} \sum_{i=1}^{t_a} (1-\alpha) I_{R_a^+,i} - \alpha I_{R_a^- - E_a^-, i}$$

and
$$R = \sum_{a \in A} \sum_{i=1}^{t_a} I_{R_a^+,i} - \beta I_{C_a,i}$$

Hence,
$$E[P] = \sum_{a \in A} t_a((1-\alpha)R_a s_a - \alpha(R_a - E_a)(1 - s_a))$$

$$E[R] = \sum_{a \in A} t_a(R_a s_a - \beta s_a)$$

Let $P^a = (1-\alpha)R_a^+ - \alpha(R_a^- - E_a^-)$ and $R_a = R_a^+ - \beta C_a$ Thus $P = \sum_{a \in A} P^a$ and $R = \sum_{a \in A} R^a$

$$\begin{aligned}\text{Var}(P^a) =& \text{Var}(\sum_{i=1}^{t_a}(1-\alpha)I_{R_a^+,i} - \alpha I_{R_a^- - E_a^-, i}) \\ =& t_a(\text{Var}((1-\alpha)I_{R_a^+,1}) + \text{Var}(-\alpha I_{R_a^- - E_a^-, 1}) \\ & + 2\text{Cov}((1-\alpha)I_{R_a^+,1}, -\alpha I_{R_a^- - E_a^-, 1})) \\ & + t_a(t_a-1)(\text{Cov}((1-\alpha)I_{R_a^+,1}, (1-\alpha)I_{R_a^+,2}) \\ & + \text{Cov}(-\alpha I_{R_a^- - E_a^-, 1}, -\alpha I_{R_a^- - E_a^-, 2}) \\ & + 2\text{Cov}((1-\alpha)I_{R_a^+,1}, -\alpha I_{R_a^- - E_a^-, 2})) \\ =& t_a((1-\alpha)^2 \text{Var}(I_{R_a^+,1}) + (\alpha)^2 \text{Var}(I_{R_a^- - E_a^-, 1}) \\ & - 2\alpha(1-\alpha)\text{Cov}(I_{R_a^+,1}, I_{R_a^- - E_a^-, 1})) \\ & + t_a(t_a-1)((1-\alpha)^2 \text{Cov}(I_{R_a^+,1}, I_{R_a^+,2}) \\ & + (\alpha)^2 \text{Cov}(I_{R_a^- - E_a^-, 1}, I_{R_a^- - E_a^-, 2}) \\ & - (1-\alpha)\alpha 2\text{Cov}(I_{R_a^+,1}, I_{R_a^- - E_a^-, 2})) \\ =& t_a((1-\alpha)^2 R_a s_a(1 - R_a s_a) \\ & + \alpha^2 (R_a - E_a)(1 - s_a)(1 - (R_a - E_a)(1 - s_a)) \\ & + 2\alpha(1-\alpha)R_a s_a(R_a - E_a)(1 - s_a)) \\ & + t_a(t_a-1)((1-\alpha)^2 R_a^2 v_a \\ & + \alpha^2 (R_a - E_a)^2 v_a + 2\alpha(1-\alpha)R_a(R_a - E_a)v_a) \\ =& t_a((1-\alpha)^2 R_a s_a(1 - R_a s_a) + \\ & \alpha^2(R_a - E_a)(1-s_a)(1 - (R_a - E_a)(1-s_a)) \\ & + 2\alpha(1-\alpha)R_a s_a(R_a - E_a)(1 - s_a)) \\ & + t_a(t_a-1)v_a(R_a - \alpha E_a)^2\end{aligned}$$

$$\begin{aligned}\text{Var}(R^a) =& \text{Var}(\sum_{i=1}^{t_a} I_{R_a^+,i} - \beta I_{C_a,i}) \\ =& t_a(\text{Var}(I_{R_a^+,1}) + \text{Var}(-\beta I_{C_a,1}) + 2\text{Cov}(I_{R_a^+,1}, -\beta I_{C_a,1})) \\ & + t_a(t_a-1)(\text{Cov}(I_{R_a^+,1}, I_{R_a^+,2}) + \text{Cov}(-\beta I_{C_a,1}, -\beta I_{C_a,2}) \\ & + 2\text{Cov}(I_{R_a^+,1}, -\beta I_{C_a,2})) \\ =& t_a(\text{Var}(I_{R_a^+,1}) + (\beta)^2 \text{Var}(I_{C_a,1}) - 2\beta\text{Cov}(I_{R_a^+,1}, I_{C_a,1})) \\ & + t_a(t_a-1)(\text{Cov}(I_{R_a^+,1}, I_{R_a^+,2}) + (\beta)^2 \text{Cov}(I_{C_a,1}, I_{C_a,2}) \\ & - 2\beta\text{Cov}(I_{R_a^+,1}, I_{C_a,2})) \\ =& t_a(R_a s_a(1 - R_a s_a) + \beta^2 s_a(1-s_a) - 2\beta R_a s_a(1-s_a)) \\ & + t_a(t_a-1)(v_a R_a^2 + v_a \beta^2 - 2\beta R_a v_a) \\ =& t_a(R_a s_a(1 - R_a s_a) + \beta^2 s_a(1-s_a) - 2\beta R_a s_a(1-s_a)) \\ & + t_a(t_a-1)v_a(R_a - \beta)^2\end{aligned}$$

Until now, we hadn't assumed anything about the correlation between $S_a$s for different $a$s. It's known that for any set of random variables $X_i$ over some values of $i$,

$$\text{Dev}(\sum_i X_i) \leq \sum_i \text{Dev}(X_i)$$

. Equality holds in the worst case where all $X_i$s are fully correlated. If $X_i$s are independent, then we have $\text{Var}(\sum_i X_i) = \sum_i \text{Var}(X_i)$. We now consider two cases.

**Unknown Correlations Case:**
If we know nothing about the correlations between $S_a$s, the best bounds we can place on $\text{Dev}(P)$ and $\text{Dev}(R)$ are $\sum_{a \in A} \text{Dev}(P^a)$ and $\sum_{a \in A} \text{Dev}(R^a)$ respectively.

When we write the constraints $E[P] \leq e_\rho \text{Dev}(P)$ and $E[P] \leq e_\rho \text{Dev}(R)$, we want the resulting optimization problem to be easy to solve. Specifically, we would like the constraints to be linear, and if that's not easily possible, then we'd like them to be convex. Expectations of $P$ and $R$ are already linear in decision variables $R_a$ and $E_a$, and so is the objective function. We will try to upper bound the rhs of both constraints (the deviations) so as to make it linear, while using a fairly tight upper bound. The rhs of both constraints are going to be a sum of standard deviations over $a \in A$. The standard deviation is a square root of the variance, which consists of a $t_a$ factor and a $t_a(t_a - 1)$ factor. The key observation is that $t_a$ is the only large quantity in the expression, where large means that is scales with the size of the table. The other terms, like $R_a$, $E_a$, $s_a$, $\alpha$, $\beta$ are all between 0 and 1. Thus only the $t_a(t_a - 1)$ factor under the square root can contribute an $O(n)$ factor to the rhs (where $n = \sum_{a \in A} t_a$ is the size of the table), while the $t_a$ factor can contribute at most $O(\sqrt{n})$. Moreover, since the cost of our final solution is likely to be $O(n)$, it is ok if the rhs is made linear by adding to it a factor that is at most $O(\sqrt{n})$.

Specifically, we have

$$\begin{aligned}
&(1-\alpha)^2 R_a s_a (1 - R_a s_a) \\
&+ \alpha^2 (R_a - E_a)(1 - s_a)(1 - (R_a - E_a)(1 - s_a)) \\
&+ 2\alpha(1-\alpha) R_a s_a (R_a - E_a)(1 - s_a) \\
&\leq 0.25(1-\alpha)^2 + 0.25\alpha^2 + 0.25 * 2\alpha(1-\alpha) \\
&= 0.25
\end{aligned}$$

$$\begin{aligned}
&R_a s_a(1 - R_a s_a) + \beta^2 s_a(1 - s_a) - 2\beta R_a s_a(1 - s_a) \\
&= (R_a s_a(1 - R_a s_a) - R_a^2 s_a(1 - s_a)) + (R_a^2 s_a(1 - s_a) \\
&\quad + \beta^2 s_a(1 - s_a) - 2\beta R_a s_a(1 - s_a)) \\
&= s_a R_a(1 - R_a) + (R_a - \beta)^2 s_a(1 - s_a) \\
&\leq s_a R_a(1 - R_a) + \max(R_a^2, (1 - R_a)^2) s_a(1 - s_a) \\
&= \max(s_a R_a(1 - R_a) + R_a^2 s_a(1 - s_a), \\
&\quad s_a R_a(1 - R_a) + (1 - R_a)^2 s_a(1 - s_a)) \\
&= \max(s_a R_a(1 - R_a + R_a - R_a s_a), \\
&\quad s_a(1 - R_a)(R_a + 1 - R_a - (1 - R_a)s_a)) \\
&= \max(s_a R_a(1 - s_a R_a), s_a(1 - R_a)(1 - (1 - R_a)s_a)) \\
&\leq \max(0.25, 0.25) \\
&= 0.25
\end{aligned}$$

And hence

$$\begin{aligned}
\text{Var}(P^a) =& t_a((1-\alpha)^2 R_a s_a(1 - R_a s_a) \\
&+ \alpha^2(R_a - E_a)(1 - s_a)(1 - (R_a - E_a)(1 - s_a)) \\
&+ 2\alpha(1-\alpha) R_a s_a(R_a - E_a)(1 - s_a)) \\
&+ t_a(t_a - 1)v_a(R_a - \alpha E_a)^2 \\
\leq & t_a^2 v_a(R_a - \alpha E_a)^2 + 0.25 t_a
\end{aligned}$$

We can then use

$$\sqrt{x^2 + y^2} \leq |x| + |y|$$

to get

$$\begin{aligned}
\text{Dev}(P^a) =& \sqrt{\text{Var}(P^a)} \\
\leq & \sqrt{v_a} t_a(R_a - \alpha E_a) + 0.5\sqrt{t_a}
\end{aligned}$$

Hence

$$\text{Dev}(P) \leq \sum_{a \in A} \sqrt{v_a} t_a(R_a - \alpha E_a) + 0.5\sqrt{t_a}$$

giving us a linear upper bound on the rhs of the constraint as needed. Similarly, we can get

$$\text{Dev}(R) \leq \sum_{a \in A} \sqrt{v_a} t_a |R_a - \beta| + 0.5\sqrt{t_a}$$

. Thus we get the convex optimization problem:

$$\text{minimize} \sum_{a \in A} t_a(R_a o_r + E_a o_e) \text{ such that}$$

$$\sum_{a \in A} (1-\alpha) t_a R_a s_a - t_a \alpha(R_a - E_a)(1 - s_a)$$
$$\geq e_\rho \sum_{a \in A} \sqrt{v_a} t_a(R_a - \alpha E_a) + 0.5\sqrt{t_a}$$
$$\sum_{a \in A} t_a R_a s_a - \beta t_a s_a \geq e_\rho \sum_{a \in A} \sqrt{v_a} t_a |R_a - \beta| + 0.5\sqrt{t_a}$$
$$R_a, E_a \in \{0, 1\}; R_a \geq E_a$$

And solving this problem gives a solution which has precision $\geq \alpha$ and recall $\geq \beta$ with probability at least $\rho$ each.

**Independent $S_a$'s case:**
Now we move to the case where $S_a$s are independent of each other. This case is worth considering because it's common. If we groups tuples by their value in a column, and evaluate a sample of tuples from each group, then we get selectivity estimates for each group, with some variance. But tuple evaluations from one group don't give us information about selectivity of other groups. Thus in case of sampling, we have independent random variables $S_a$ for each $a \in A$.

Because of the independence, we have

$$\text{Var}(P) = \sum_{a \in A} \text{Var}(P^a)$$

and

$$\text{Var}(R) = \sum_{a \in A} \text{Var}(R^a)$$

Thus in the optimization problem constraints are

$$E[P] \leq e_\rho \sqrt{\sum_{a \in A} \text{Var}(P^a)}$$

*
$$E[R] \leq e_\rho \sqrt{\sum_{a \in A} \text{Var}(R^a)}$$

Again, the objective function as well as lhs of both constraints are linear in decision variables $R_a$ and $E_a$. The rhs has a square root. Like we did before, we would like to upper bound the rhs so as to make the optimization problem linear or convex. In this case, there does not seem to be a way to make the rhs linear without increasing it significantly. But it is possible to make the problem convex instead, with a slight increase in the rhs.

As in the last section, we note that

$$\text{Var}(P^a) \leq t_a^2 v_a (R_a - \alpha E_a)^2 + 0.25 t_a$$

Thus

$$\text{Dev}(P) \leq \sqrt{\sum_{a \in A} t_a^2 v_a (R_a - \alpha E_a)^2 + 0.25 t_a}$$

The rhs is a square root of a sum of squares and a positive constant. Suppose we have a $|A| + 1$ dimensional vector $\vec{R}$ whose components are the $R_a \sqrt{t_a^2 v_a}$ for each $a \in A$ and the constant $\sqrt{\sum_{a \in A} 0.25 t_a}$ and a $|A| + 1$ dimensional vector $\vec{E}$ whose components are $\alpha E_a \sqrt{t_a^2 v_a}$ for each $a \in A$ and 0. Then the rhs is given by $\left\| \vec{R} - \vec{E} \right\|_2$. For vectors $\vec{R}_1$, $\vec{E}_1$, $\vec{R}_2$, $\vec{E}_2$, by the triangle inequality, we have

$$\left\| \vec{R}_1 - \vec{E}_1 \right\|_2 + \left\| \vec{R}_2 - \vec{E}_2 \right\|_2 \geq 2 \left\| \frac{\vec{R}_1 - \vec{E}_1}{2} + \frac{\vec{R}_2 - \vec{E}_2}{2} \right\|_2$$

$$= 2 \left\| \frac{\vec{R}_1 + \vec{R}_2}{2} - \frac{\vec{E}_1 + \vec{E}_2}{2} \right\|_2$$

Thus the function $\left\| \vec{R} - \vec{E} \right\|_2$ is convex over the $R_a$s and $E_a$s.

Now consider the precision constraint

$$E[P] \geq e_\rho \left\| \vec{R} - \vec{E} \right\|_2$$

Since the lhs is linear in $R_a$s and $E_a$s, while the rhs is convex, the constraint is convex. Similarly, we can show that upper bounding $\text{Dev}(R)$ by $\sqrt{\sum_{a \in A} t_a^2 v_a (R_a - \beta)^2 + 0.25 t_a}$ makes the recall constraint convex. Thus we reduce our problem to the following convex optimization problem

$$\text{minimize} \sum_{a \in A} t_a (R_a o_r + E_a o_e) \text{ such that}$$

$$\sum_{a \in A} (1 - \alpha) t_a R_a s_a - t_a \alpha (R_a - E_a)(1 - s_a)$$

$$\geq e_\rho \sqrt{\sum_{a \in A} t_a^2 v_a (R_a - \alpha E_a)^2 + 0.25 t_a}$$

$$\sum_{a \in A} t_a R_a s_a - \beta t_a s_a \geq e_\rho \sqrt{\sum_{a \in A} t_a^2 v_a (R_a - \beta)^2 + 0.25 t_a}$$

$$R_a, E_a \in \{0, 1\}; R_a \geq E_a$$

which can be solved with any local optimization technique.

### 10.5 Tightness Proof for Estimated Selectivities Case

### 10.6 Justification for Rule of Thumb from Section 4

We will justify our rule of thumb, by making some simplifying assumptions in our equations, and then optimizing our problem locally. To start with, we ignore the exact values of $R_a$s, $E_a$s and $s_a$s and instead write the precision and recall constraints approximately as

$$\sum_{a \in A} F_a \tau_1 - \sqrt{\sum_{a \in A} \frac{\tau_2 t_a^2}{F_a}} \geq 0$$

for some constants $\tau_1$ and $\tau_2$.

Then, the derivative of the lhs with respect to $F_a$ is:

$$\tau_1 - \frac{t_a^2 \tau_2}{2 F_a^2 \sqrt{\sum_{a \in A} \frac{\tau_2 t_a^2}{F_a}}}$$

The derivative is negative when $F_a \approx 0$, and thus increasing $F_a$ helps satisfy the constraint better (compared to increasing other $F$s or $E$s with larger derivative). As $F_a$ increases, the derivative decreases, and the marginal value of increasing $F_a$ decreases. In the locally optimal solution, the derivative with respect to each $F_a$ will be equal. Suppose for the locally optimal solution,

$$Y = 2 \sqrt{\sum_{a \in A} \frac{\tau_2 t_a^2}{F_a}}$$

, and the derivative with respect to each $F_a$ is $\tau_3$. Then, we have

$$\tau_1 - \tau_3 = \frac{t_a^2}{F_a^2 Y}$$

and hence

$$F_a \approx \frac{t_a}{\sqrt{Y(\tau_1 - \tau_3)}}$$

Let $n = \sum_{a \in A} t_a$ be the total number of tuples in the table. Now we want to find the order of magnitude value of $Y$. We have

$$\frac{t_a}{F_a} = \sqrt{Y((\tau_1 - \tau_3))}$$

and thus

$$Y = 2 \sqrt{\sum_{a \in A} \frac{\tau_2 t_a^2}{F_a}}$$

$$= 2 \sqrt{\sum_{a \in A} \sqrt{Y(\tau_1 - \tau_3)} \tau_2 t_a}$$

$$\Rightarrow Y^{\frac{3}{4}} \approx 2 \sqrt{\tau_2 \sqrt{(\tau_1 - \tau_3)} n}$$

$$\Rightarrow Y = (4 \tau_2 \sqrt{\tau_1 - \tau_3} n)^{\frac{2}{3}}$$

$$\Rightarrow F_a = \frac{t_a}{\sqrt{(4 \tau_2 \sqrt{\tau_1 - \tau_3} n)^{\frac{2}{3}} (\tau_1 - \tau_3)}}$$

$$\Rightarrow F_a \approx \tau_4 t_a n^{-\frac{1}{3}}$$

where $tau_4$ is a constant in terms of the other $\tau$s. This justifies our rule of thumb, with the constant $\tau_4$ being the num parameter in the rule of thumb.

### 10.7 Extensions

*10.7.1 Alternate Objective Functions*

Our paper focuses on the scenario where precision and recall constraints are specified, while we try to minimize cost. An interesting variation of the problem is where a user has a fixed cost budget, and wishes to maximize the number correct tuples returned while maintaining some level of precision and not exceeding the budget.

Minor modifications of our techniques can be used to handle this variation. The cost now becomes one of the constraints, while recall (or precision) becomes the objective function to be maximized. We can tighten the constraints slightly to turn the resulting problem into a convex optimization problem, and solve it to maximize precision or recall while keeping cost within a given limit.

### 10.7.2 Multiple Predicates

Another variation of the problem is where the query might have multiple chained selects on a table (which is equivalent to a conjunction of multiple predicates). If any of these predicates is not an expensive UDF, then it makes sense to execute those predicate first. But after doing this, we may still have more than one UDF in the select condition, which creates an interesting variation of our problem. The precision and recall constraint are specified by the user for the final output, making it possible to trade-off accuracy in one predicate for higher accuracy in the other at the same cost. Also, the probability of a tuple being correct is now affected by the probabilities of it satisfying both the predicates, so if a tuple that is very unlikely to satisfy one predicate, then we may not bother to evaluate the second predicate on it even if it was likely to satisfy the second predicate (because it probably won't be in the final output anyway).

We can extend our method to solve this variation of the problem, by introducing decision variables representing combined decisions on the predicates (For example, there may be a decision variable $RR_a$ for group $a$ that is 1 if and only if we are assuming both predicates are true for the tuple). This gives us a problem where the number of variables is exponential in the number of predicates, but still linear in table size (which is likely to be much higher than number of predicates anyway), which we can solve with a variation of our techniques.

### 10.7.3 Single Predicate with Join

Another variation we can consider is where the table $T$ being selected on is later going to be joined with another table $T_2$. Each tuple of the table $T$ may match with a different number of tuples of $T_2$. Thus it may be worthwhile for us to evaluate a tuple with low correctness-probability that matches with a large number of tuples from $T_2$, over a tuple with higher correctness probability that joins with fewer tuples from $T_2$.

We can solve this variation by
- Creating a separate decision variable for each value in the join column. (Thus if $a \in A$ is a value in the correlated column, and $j$ is a value in the join column, then we have decision variables $R_{a,j}$ and $E_{a,j}$).
- Multiplying the $R_{a,j}$ and $E_{a,j}$ terms in the precision and recall constraints by the number $n_j$ of tuples from $T_2$ that have value $j$ in the join column.

## 10.8 Additional statistics on our datasets

Table 3 shows further describes our datasets and the groups formed using the correlated column chosen by our scheme described in Section 4.4. For each dataset, we describe the number of groups, the Standard Deviation of group size, Standard Deviation of group selectivity, and the Pearson correlation coefficient between size and selectivity.

| Dataset | Num. Groups | Size Dev. | Selectivity Dev. | Correlation |
|---|---|---|---|---|
| LC | 7 | 5233 | 0.13 | 0.84 |
| Prosper | 8 | 1521 | 0.20 | 0.20 |
| Census | 7 | 8183 | 0.15 | 0.36 |
| Marketing | 10 | 5070 | 0.20 | $-0.65$ |

*Table 3: Size, Selectivity statistics for Dataset Groups*